\documentclass{egpubl}
\usepackage{eurovis2026}

\SpecialIssuePaper         % uncomment for final version of Computer Graphics Forum, special issue
\CGFccby
\usepackage[T1]{fontenc}
\usepackage{dfadobe}  

\usepackage{cite}  % comment out for biblatex with backend=biber
\BibtexOrBiblatex
\electronicVersion
\PrintedOrElectronic
\ifpdf \usepackage[pdftex]{graphicx} \pdfcompresslevel=9
\else \usepackage[dvips]{graphicx} \fi

\usepackage{egweblnk}
\usepackage{xspace}
\usepackage{tikz}
\usepackage{booktabs}
\usepackage{tcolorbox}
\usepackage{tabularx}
\usepackage{array} % for \newcolumntype
\usepackage{listings}
\tcbuselibrary{skins,breakable,listings}
\usepackage[table]{xcolor}
\usepackage[table]{xcolor}
\usepackage{pgfmath} % for \pgfmathsetmacro

\newcolumntype{Y}{>{\raggedright\arraybackslash}X} % 左对齐可换行 X 列
\newcolumntype{P}[1]{>{\raggedright\arraybackslash}m{#1}}
\newcolumntype{C}[1]{>{\centering\arraybackslash}m{#1}}
\newcommand{\name}{\textit{ReasonDiag}\xspace}
\newcommand{\Overview}{Overview\xspace}
\newcommand{\SecView}{Section View\xspace}
\newcommand{\OriView}{Original CoT\xspace}

\definecolor{CaptionColor}{RGB}{89, 91, 97}
\newcommand*\captionID[1]{\tikz[baseline=(char.base)]{
            \node[shape=rectangle,fill=CaptionColor, text=white, inner sep= 1pt,minimum size=8pt,rounded corners=1pt] (char) {\textbf{#1}}}}

\definecolor{pos}{RGB}{0,48,87}     
\definecolor{neg}{RGB}{128,0,32}    
\newcommand{\DeltaBox}[2]{
  \pgfmathsetmacro{\val}{#1}
  \pgfmathsetmacro{\maxv}{1}
  \pgfmathsetmacro{\pct}{min(100,abs(\val)/\maxv*100)}
  \ifdim \val pt > 0pt
    \colorbox{pos!\pct!white}{\strut\,#2\,}
  \else\ifdim \val pt < 0pt
    \colorbox{neg!\pct!white}{\strut\,#2\,}
  \else
    \colorbox{white}{\strut\,#2\,}
  \fi\fi
}

\newcommand{\csw}[1]{\textcolor{black}{#1}}

\newcommand{\note}[1]{\textcolor{black}{#1}}

\title[Interactive Diagnosis of LLM Chain-of-Thought Errors]%
      {When the Chain Breaks: Interactive Diagnosis of LLM Chain-of-Thought Reasoning Errors}

\author[S. Chen \& N. Sritharan \& X. Wen \& C. Zhang \& X. Wang \& Y. Wang]
{\parbox{\textwidth}{\centering 
Shiwei Chen$^{1}$\orcid{0009-0007-9106-7989}, 
Niruthikka Sritharan $^{1}$\orcid{0009-0007-3322-3218}, 
Xiaolin Wen $^{1}$\orcid{0000-0002-8562-7640},  
Chenxi Zhang $^{1}$ \orcid{0009-0005-3814-8443}, 
Xingbo Wang $^{2}$\orcid{0000-0001-5693-1128}, 
Yong Wang $^{1}$\thanks{Corresponding author: yong-wang@ntu.edu.sg}\orcid{0000-0002-0092-0793}
}
        \\
{\parbox{\textwidth}{\centering 
        $^1$ College of Computing and Data Science, Nanyang Technological University, Singapore\\
        $^2$ Bosch Research North America, Sunnyvale, CA, United States
       }
}\\
}

\begin{document}

% uncomment for using teaser
% \teaser{
%  \includegraphics[width=0.9\linewidth]{eg_new}
%  \centering
%   \caption{New EG Logo}
% \label{fig:teaser}
%}

\maketitle
%-------------------------------------------------------
\begin{abstract} 

    % \wy{Should we add the keyword ``reasoning'' to the paper title? i.e., ``When the Chain Breaks: Interactive Diagnosis of LLM Chain-of-Thought Reasoning Errors''}

   Current Large Language Models (LLMs), especially Large Reasoning Models, can generate Chain-of-Thought (CoT) reasoning traces to illustrate how they produce final outputs, thereby facilitating trust calibration for users.
   However, these CoT reasoning traces are usually lengthy and tedious, and can contain various issues, such as logical and factual errors, which make it difficult for users to interpret the reasoning traces efficiently and accurately. 
   To address these challenges, we develop an error detection pipeline that combines external fact-checking with symbolic formal logical validation to identify errors at the step level.
   Building on this pipeline, we propose \name, an interactive visualization system for diagnosing CoT reasoning traces. 
   \name provides 1) an integrated arc diagram to show reasoning-step distributions and error-propagation patterns, and 2) a hierarchical node-link diagram to visualize high-level reasoning flows and premise dependencies. 
    We evaluate \name through a technical evaluation for the error detection pipeline,
    two case studies,
    and user interviews with 16 participants. The results indicate that \name helps users effectively understand CoT reasoning traces, identify erroneous steps, and determine their root causes.

\begin{CCSXML}
<ccs2012>
   <concept>
       <concept_id>10003120.10003145.10003147.10010365</concept_id>
       <concept_desc>Human-centered computing~Visual analytics</concept_desc>
       <concept_significance>500</concept_significance>
       </concept>
   <concept>
       <concept_id>10003120.10003121.10003129</concept_id>
       <concept_desc>Human-centered computing~Interactive systems and tools</concept_desc>
       <concept_significance>300</concept_significance>
       </concept>
 </ccs2012>
\end{CCSXML}

\ccsdesc[500]{Human-centered computing~Visual analytics}
\ccsdesc[500]{Human-centered computing~Interactive systems and tools}

\printccsdesc   
\end{abstract}  
%--------
\section{Introduction}
% \wy{Pls update the keywords in the abstract.}

% % 背景：人们使用在错误敏感task中使用 LLM -> cot提供了人们monitor llm 的 unique opportunity. 
% Nowadays, people increasingly depend on large language models (LLMs) in tasks where errors carry real-world consequences, such as information seeking and decision support~\cite{chatterji2025people}. Yet as probabilistic black boxes, LLMs offer no inherent mechanisms to justify their outputs or reveal their reasoning pathways~\cite{barez2025chain}. 
% Chain-of-thought (CoT) reasoning in large reasoning models (LRMs) has emerged as a critical solution:
% % With Chain-of-Thought (CoT) reasoning, large reasoning models (LRMs) have emerged as a critical solution: 
% beyond improving task performance~\cite{wei2022chain}, it generates inspectable reasoning traces that offer a unique opportunity to monitor LLM behavior~\cite{korbak2025chain}.

In the past few years, large language models (LLMs) have become foundational across diverse real-world tasks, such as information seeking, decision-making, and practical guidance~\cite{chatterji2025people}.
% Despite their fast-growing popularity and significant importance, LLMs are intrinsically probabilistic black-box models and lack transparency.
% \wy{1. pls double check it. 2. For the final revised draft, please use red, instead of blue, to highlight the revised parts, as blue is also used for citations.}
\csw{
Chain-of-Thought (CoT) prompting~\cite{wei2022chain} improves task performance by encouraging models to output intermediate reasoning steps in natural language, producing a human-readable reasoning trace. 
% Chain-of-Thought (CoT) prompting~\cite{wei2022chain} has become a widely-used technique to improve LLMs' task performance, and it encourages LLMs to output intermediate reasoning steps in natural language (i.e., \textit{CoT reasoning traces}), which can help LLM users understand how LLMs arrive at their answer.
% More specifically, large reasoning models (LRMs) are a type of LLMs with an emphasis on multi-step reasoning, often enabling the generation of reasoning traces. Many LRMs (though not all of them) can produce CoT reasoning traces in their outputs when performing reasoning tasks~\cite{team2023gemini, Claude, guo2025deepseek,qwen}\footnote{See Appendix~\ref{app:api-exposure}, Table~\ref{tab:api-reasoning-exposure} for a summary of how some major LRM providers expose reasoning traces through user interfaces and APIs.}.  
%
%
Building on this idea, large reasoning models (LRMs) go further by producing reasoning traces natively during inference~\cite{openai_api,team2023gemini, guo2025deepseek,qwen}\footnote{See Appendix~\ref{app:api-exposure} for a summary of how some major LRM providers expose reasoning traces through user interfaces and APIs.}. 
Apart from performance improvements, these reasoning traces offer user-centric value: they enable early detection of model misbehavior~\cite{baker2025monitor,guan2025monitoring, korbak2025chain}, facilitate the calibration of user trust by revealing the model reasoning~\cite{barez2025chain, wei2022chain}, and can even serve as an interactive medium to steer model output~\cite{pang2025interactive}. }

% Chain-of-Thought (CoT) reasoning~\cite{wei2022chain} provides a step-by-step explanation of the decision process of LLMs, especially Large Reasoning Models (LRMs), making their reasoning more interpretable. Besides being applied to improve LLM task performance~\cite {wei2022chain}, 
% CoT reasoning can also be used to generate explanations for LLM users to monitor how LLMs generate final outputs~\cite{korbak2025chain}. CoT reasoning has been integrated into many mainstream LLMs such as DeepSeek‑R1~\cite{guo2025deepseek}, GPT-5~\cite{chatgpt2023}, and Gemini 3~\cite{team2023gemini}. 

% 但存在的 challenges 是 1. text 长；2. 很可能存在错误；
However, the utility of CoT reasoning traces is often undermined by two primary challenges: \textbf{\textit{verbosity}} and \textbf{\textit{possible unreliability}}. 
First, \csw{long CoT reasoning traces have become a common feature of
% modern reasoning models
most LRMs~\cite{chen2025towards}. These traces are characterized by a large set of reasoning steps that explore diverse paths as well as steps that revisit earlier decisions~\cite{chen2025towards}. Consequently, the verbose reasoning traces and their non-linear structures make it time-consuming for common users to understand the reasoning process and mentally reconstruct the complex dependencies among various exploratory claims or revised statements.}
Second, due to the lack of internal verification mechanisms, LRMs often generate reasoning traces that suffer from logical and factual errors~\cite{tyen-etal-2024-llms}.
Such errors are often distributed in different parts of the lengthy CoT reasoning traces, and the verification of a statement also depends on the context and premise statements, making it difficult to identify reasoning errors and often resulting in users' misplaced trust in the model output. 

% 当前的方法独立解决两个challenge
Limited research has been done on the interpretation and diagnosis of CoT reasoning traces.
Existing LLM interfaces (e.g., ChatGPT~\cite{chatgpt2023}) often summarize or selectively hide intermediate reasoning steps, 
% \wy{I do not know which part of ChatGPT you are referring to}
which strips away critical context and undermines the verifiability of the model reasoning process.
More recent visualization approaches, like HIPPO~\cite{pang2025interactive}, iGraph~\cite{zhou2025improving} and ReasonGraph~\cite{li-etal-2025-reasongraph}, provide a graph-based organization of LLM reasoning traces. But they do not enable diagnostic support, and users still need to fully rely on the original CoT reasoning traces to understand and diagnose errors. 
The research in the AI field has attempted to reduce the errors in the LLM CoT reasoning traces via self-consistency checking~\cite{wangself} and 
% process reward models (PRMs)
process rewarding~\cite{lightman2023let}. However, these techniques can only reduce errors to some extent in CoT reasoning traces, which makes tedious human examination still necessary.
Therefore, an effective method to efficiently understand and diagnose CoT reasoning traces is still missing and urgently needed.

\begin{figure*}
    \centering
    \includegraphics[width=1\textwidth]{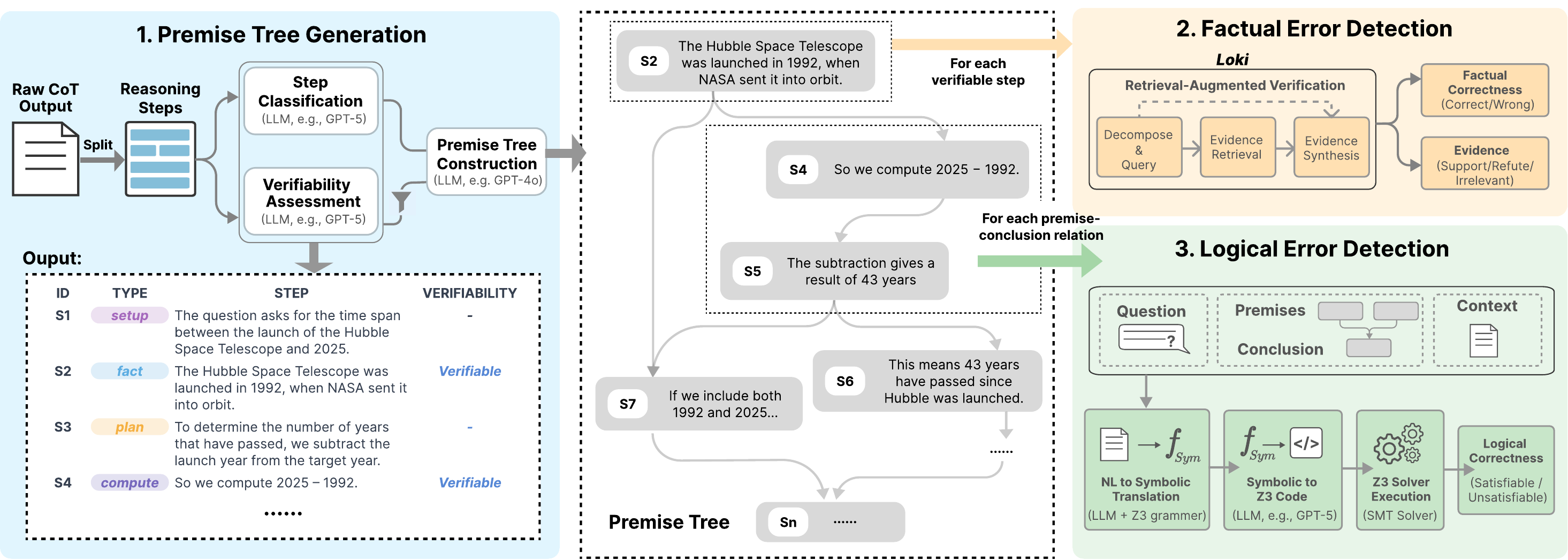}
    \caption{The Error Detection Pipeline comprises three stages: (1) \textbf{Premise Tree Generation} structures the raw CoT by classifying step roles and mapping dependencies; (2) \textbf{Factual Error Detection} verifies checkable claims using retrieval-augmented external evidence; and (3) \textbf{Logical Error Detection} translates natural language steps into symbolic constraints for formal consistency checks via the Z3 solver. \csw{The CoT reasoning example above is prompted by the question: ``How many years have passed between the launch of the Hubble Space Telescope and the year 2025?''}}
    \vspace{-5mm}
    \label{fig:error-detection-pipeline}
\end{figure*}

% 我们的方法从结合两种，通过 fact check 结合逻辑和用户对 cot的理解。
To fill the research gap, we propose \name, an interactive visual analytics approach to help LLM users effectively interpret and diagnose \csw{CoT reasoning traces}. 
%
%
% \csw{Specifically, instead of autonomous correction, \name focuses on human-in-the-loop diagnosis of factual errors and logical errors, with the ability to support complex reasoning traces ranging from dozens to over a hundred reasoning steps.}
\csw{Specifically, \name focuses on human-in-the-loop diagnosis of factual and logical errors in CoT reasoning traces, instead of error correction. Our system targets general LRM users who rely on model outputs and need to review CoT reasoning traces for trust calibration. It aims to help LRM users explore individual complex reasoning traces ranging from dozens to over a hundred reasoning steps.}
% \wy{What is your motivation for adding this sentence here? Please check the revised draft.}
%
%
\name is designed based on the requirements distilled from a formative study with nine experienced LRM users, which examined their challenges and strategies when diagnosing CoT reasoning traces.
\name integrates an automated error detection pipeline with interactive visualizations.
The error detection pipeline combines two complementary verification paradigms: external fact-checking and formal logic verification, which collectively address the challenge of possible unreliability in CoT reasoning traces. 
External fact-checking is invoked for claims that rely on real-world facts, domain-specific knowledge, or commonsense reasoning that cannot be captured by formal logic alone (e.g., the claim \textit{``The Hubble Space Telescope was launched in 1992''} shown in~\autoref{fig:error-detection-pipeline}). 
Formal logic verification operates on reasoning steps
and checks whether each step logically follows its stated premises (e.g., we verify whether the step \textit{``So we compute 2025-1992''} in ~\autoref{fig:error-detection-pipeline} is really warranted by the premise \textit{``The Hubble Space Telescope was launched in 1992''}).

Two coordinated visualizations are developed to enable effective exploration and diagnosis of CoT reasoning traces: an arc diagram provides an overview of error propagation patterns across the chain (\autoref{fig:interface}\captionID{A}), and a hierarchical node-link diagram exposes the high-level reasoning structure and premise–conclusion dependencies (\autoref{fig:interface}\captionID{B}), where the original CoT reasoning traces can also be revealed on demand to show the reasoning context.
This design allows users to rapidly navigate, filter, trace CoT reasoning dependencies, and understand root causes of errors while preserving the full contextual nuance and enabling human verification.

We evaluate \name through a technical evaluation of our error-detection pipeline, two case studies, and user interviews with 16 participants. The results show that \name effectively helps users comprehend long CoT reasoning traces, pinpoint erroneous steps, and diagnose their root causes compared to reading plain texts alone. 
In summary, our main contributions are as follows:       
\begin{itemize}
    \item A formative study that systematically characterizes the limitations of raw textual CoT for human oversight;
    \item An interactive visualization system named \name, which integrates an error-detection pipeline with coordinated arc and node-link views to support scalable and interactive exploration and diagnosis of CoT reasoning traces;
    \item A technical evaluation of the error-detection pipeline, two case studies, and user interviews with 16 participants to demonstrate the usefulness and effectiveness of \name.
\end{itemize}

% \wy{LLMs vs. LRMs: both terms are widely used in the paper. It may be better to stick to one term. What will be our final decision here?}
\section{Related Work}

This work is related to prior research on text visualization for LLMs and verification of CoT reasoning.

\textbf{Text Visualization for LLMs}.
% often struggle with long reasoning traces due to scalability issues
Text visualization is increasingly used for LLMs for their text-in, text-out nature. Prior work on text visualization for LLMs can be grouped into graph-based, text-centric, heatmap-based, spatial, and multi-view approaches.
Graph-based representations are widely used; systems like Graphologue~\cite{jiang2023graphologue} and WaitGPT~\cite{xie2024waitgpt} improve structural transparency by converting long text into node-link graphs. More recent works like HIPPO~\cite{pang2025interactive}, ReasonGraph~\cite{li-etal-2025-reasongraph}, and iGraph~\cite{zhou2025improving} also leverage graphs to visualize reasoning traces. Text-centric visualizations, such as in-context highlighting~\cite{lee2025llm}, are also commonly used for being intuitive and preserving the original narrative structure. 
Heatmap-based visualizations, like LLM Analyzer~\cite{cheng2025understanding}, help users quickly spot important text spans. Spatial visualizations like Landscape of Thoughts~\cite{zhou2025landscape} embed intermediate reasoning states into a 2D space, helping users spot anomalous patterns. Multi-view systems like CommonsenseVIS~\cite{wang2023commonsensevis} combine coordinated views to support flexible exploration. \csw{
We adopt two coordinated visualizations, i.e., 
an arc diagram and a hierarchical node-link diagram, to support rapid navigation, filtering, and dependency tracing of reasoning traces. Furthermore, inspired by systems like Pluto~\cite{pluto}, we enable bidirectional interactions to align the raw text with these visual structures, enabling a seamless analytical experience.}
% \wy{Why adding this?}

\textbf{Verification of CoT Reasoning}.
% 1.Most existing work uses first-order logic (FOL) to generate reasoning chains, rather than to analyze an already produced reasoning chain; 
% 2.for the work"Premise-Augmented Reasoning Chains for Enhanced Error Identification in Large Language Models" which use fol to anaylsis cot. it does not take into account sentences that cannot be naturally represented in FOL, and some facts cannot be verified by FOL.
% Therefore, we introduce a human in the loop: while FOL is used to verify the logical correctness of the reasoning/fact check are used for xxx, and also human users help quickly interpret and assess where and how the CoT goes wrong.
Recent studies on verifying CoT reasoning have made progress by integrating First-Order Logic (FOL) and other symbolic methods into LLM reasoning. One line of work, such as LOGIC-LM \cite{pan2023logic} and SymbCoT \cite{xu2024faithful}, converts problems into FOL and performs symbolic deduction to generate faithful logical reasoning. Another smaller line of work, e.g., SWRV~\cite{chen2025llmsr}, analyzes existing CoTs by using SMT solvers to check individual steps. However, these approaches still rely on custom symbolic representations and cannot fully audit natural-language reasoning, especially for implicit assumptions or steps outside formal logic.
PARC \cite{mukherjee2025premise} structures CoTs into premise-linked graphs for fine-grained error detection, but it lacks a formal FOL representation and cannot handle reasoning requiring domain knowledge, commonsense, or non-formalizable facts. 
% Human-in-the-loop verification
\csw{Different from existing work, we propose a human-in-the-loop verification framework that integrates symbolic reasoning, fact-checking, and user review. Following the prior iterative human-AI workflow where users inspect and correct intermediate outputs~\cite{wang2025data}, we treat automatic verification as suggestions supported by evidence, and route ambiguous or likely-wrong cases to human validation. This hybrid approach preserves symbolic precision while improving reliability through human validation of ambiguous cases and extending its coverage beyond FOL.}

% To address these limitations, we propose a human-in-the-loop verification framework combining symbolic reasoning, fact-checking, and human review. FOL is applied to steps that can be formally represented, external fact-checking covers knowledge beyond formal logic, and human reviewers ensure interpretability and catch unsupported assumptions. This hybrid approach preserves symbolic precision while expanding coverage to reasoning steps not captured by FOL alone.

\section{Formative Study}
 To inform the design of our error detection pipeline, we conducted a formative study with 9 experienced LRM users to
 % \csw{validate the demand} 
 \csw{confirm their needs of inspecting CoT reasoning traces}
 and understand their current workflows, strategies, and pain points when diagnosing errors in CoT reasoning traces.
 % \wy{LRM or LLMs? We are using both terms in the paper, which is a bit confusing. Can we stick to only one term as much as possible throughout the paper?}

\subsection{Participants and Procedure}
\csw{
Our system targets general LRM users who inspect reasoning traces of LRMs for better trust calibration. However, because 
mainstream LRM products (e.g., ChatGPT~\cite{chatgpt2023}) often abstract away these intermediate steps, general users currently lack established workflows to effectively inspect CoT for trust calibration. 
To identify the bottlenecks of trace inspection and derive user requirements, we purposefully recruited 9 experienced users (age 21–31, Avg = 25.2) with extensive familiarity with LRMs (M = 6.7 on a 7-point scale). To validate the practical need for this analytical task,  participants confirmed their need to review reasoning traces, with eight out of nine indicating their frequency of reviewing reasoning traces as \textit{Sometimes} or \textit{Often}. More details can be found in Appendix~\ref{appendix:formative_study}.}

The study was conducted via an online survey. After providing demographic information and reporting LRM familiarity, participants reviewed an example CoT containing errors and performed three tasks: understanding the CoT reasoning trace, identifying erroneous steps, and determining error root causes. Survey items included scaled ratings, multiple-selection questions, and open-ended questions to capture both quantitative and qualitative insights into participants’ reasoning strategies, challenges, and perceptions of features that could support error analysis. 

\subsubsection{Findings}
Analysis of participant responses yielded four main findings regarding challenges and strategies in diagnosing reasoning traces: 

\begin{itemize}
    \item \textbf{F1. Users Prioritize Different Reasoning Steps.} 
    Participants focused on distinct parts of the reasoning process: some examined the initial plan (\textit{``first think whether LLM’s plan is reasonable,''} P2), others centered on key assumptions or results (\textit{``where key assumptions or results appear,''} P9), and some monitored signals of model difficulty such as repetition (P1: \textit{``look for errors in the repetitive thinking process''}).
    
    \item \textbf{F2. Long CoTs Induce Cognitive Overload.}
    Many described CoTs as overly long, repetitive, and poorly structured, making it difficult to form a coherent mental model (\textit{``The CoT is too long / No summary,''} P1, P3, P5). Participants noted that verbosity also degraded coherence; for example, P6 observed that the model \textit{``usually forgets the operational logic used earlier.''} They recommended more concise formats with highlighted assumptions and turning points.
    
    \item \textbf{F3. Retroactive or Nonlinear Logic Reduces Traceability.}
    Participants struggled to connect premises to conclusions when reasoning contained self-corrections, branching, or retroactive edits (\textit{``I couldn’t trace premises to conclusions,''} P2, P4, P5, P7). Abrupt transitions further weakened interpretability (\textit{``some reasoning steps make abrupt jumps and lack tight connections to earlier statements''}, P7).
    
    \item \textbf{F4. Limited Verification Resources Hinder Error Detection.}
    Participants relied on external checks (\textit{``verify the reasoning step by searching via the browser,''} P4; \textit{``compare with my own knowledge and external information,''} P7) and expressed a need for automated consistency checking (P9: a tool that \textit{``reviews each step and flags possible errors''}).

\end{itemize}

\subsection{Design Requirements}
Based on the findings from our formative study and insights from related work\csw{~\cite{mukherjee2025premise, bogdan2025thought, SHNEIDERMAN2003364}}, we derive the following design requirements to support effective diagnosis of errors in CoT reasoning traces:

\begin{itemize}
    \item \textbf{R1. Enable automated detection of potential errors.}  
    % To address participants' need for external verification methods (F4), the system should automatically detect potential errors in reasoning traces before manual inspection.
    \csw{A possible solution to address participants' need for external verification methods (F4) is to automatically detect potential errors in reasoning traces before manual inspection~\cite{mukherjee2025premise}.}
    This will help users quickly identify suspicious steps that need examination, thereby facilitating efficient error diagnosis workflows.
    \item \textbf{R2. Provide an overview of the overall reasoning structure.} An overview is crucial for managing lengthy CoT outputs (F2) and accommodating diverse diagnostic strategies (F1)\csw{~\cite{SHNEIDERMAN2003364}}. The system should present high-level summaries and visual representations of the reasoning flow, enabling users to rapidly grasp the global reasoning flow and navigate to desired sections without the need to process verbose, repetitive details.
    \item \textbf{R3. Support traceable navigation across individual reasoning steps.} Our formative study indicated that retroactive logic structures cause traceability issues (F3), hindering both CoT understanding and error identification. \csw{A practical approach is to allow users to trace backward to the supporting premises of a step as well as connect it to its dependent conclusions~\cite{bogdan2025thought}. Therefore, the system should enable traceable navigation for each individual step.}
    \item \textbf{R4. Support identifying errors in CoT and analyzing their propagation.} Addressing the need to understand how errors influence subsequent reasoning (F3) and the requirement for a visual hint for possible errors (F4), the system should allow users to pinpoint where errors occur and visualize how these errors influence subsequent reasoning steps.
    \item \textbf{R5. Enable in-depth cause analysis of individual errors.} To address users' need for in-depth cause analysis (F4), the system should provide mechanisms for investigating why specific steps are incorrect. This is crucial for two reasons. First, since no automated error detection method is perfect, users need to verify whether flagged errors are supported by valid evidence or are false positives. Second, understanding error causes is essential for effective correction.
\end{itemize}

\section{Error Detection Pipeline} \label{sec-error-dec-pipeline}
Guided by the need to enable automated detection of potential errors (R1), we propose an error detection pipeline that combines external fact-checking with symbolic formal logical validation to identify both factual and logical errors at the step level.

\subsection{Premise Tree Generation}
This stage processes the raw CoT output to identify the functional role of each reasoning step and establish premise-conclusion relations. The stage consists of four processes: (1) splitting the reasoning trace into individual steps, (2) classifying each reasoning step by its functional role, (3) verifying the verifiability of each step, and (4) constructing a premise tree that explicitly captures premise-conclusion relations between verifiable steps. \csw{The details of prompts and model parameters can be found in Appendix~\ref{appendix:prompting_details}.}

\textbf{Step Segmentation.} Following the step-wise evaluation paradigm used in process reward models (PRMs)~\cite{lightman2023let}, we segment the reasoning trace into individual sentences, each representing a reasoning step.

\textbf{Step Classification.} Each reasoning step is then classified to support users' focus on different types of reasoning steps. The classification adopts the taxonomy and prompting strategy from ThoughtAnchor~\cite{bogdan2025thought}, which is grounded in the findings that reasoning steps exhibit distinct reasoning functions~\cite{venhoffunderstanding}. Specifically, an LLM-based classifier categorizes reasoning steps by their functional roles, including: problem setup, plan generation, fact retrieval, active computation, uncertainty management, result consolidation, self-checking, and final answer emission. This taxonomy is adopted because it provides comprehensive coverage of reasoning functions in CoT reasoning traces, aligning with users' diverse diagnostic strategies observed in the formative study. 

\textbf{Verifiability Assessment.} Alongside classification, we assess whether each step is verifiable. This is necessary because many steps contain no checkable claim and instead serve organizational or procedural roles~\cite{wei2022chain}. By filtering for verifiable steps, the system can focus verification resources on meaningful steps, improving efficiency and avoiding unreliable judgments. A step is classified as verifiable if it contains an objectively checkable claim (e.g., external fact: \textit{``Paris is the capital of France''}; or logical validity: \textit{``If A > B and B > C, then A > C''}). Conversely, a step is classified as non-verifiable if it only consists of (e.g., procedural directives: \textit{``Let us consider this approach''}; reflective commentary: \textit{``This seems reasonable''}). Recent work demonstrates that LLMs achieve high accuracy on verifiability classification~\cite{majer2024claim}. We therefore adopted an LLM-based classifier with few-shot prompting to ensure robust classification.

\textbf{Premise Tree Construction.} After classification and verifiability assessment, we construct a premise tree that explicitly encodes premise--conclusion relationships among verifiable steps. By transforming a linear reasoning trace into a structured reasoning flow, the premise tree enables downstream logical error detection and error propagation analysis (R3, R4). To construct the premise tree, we adopt the aggregative premise mapping method from PARC~\cite{mukherjee2025premise}, as their study demonstrates that LLMs are well-suited for this task, achieving 90\% recall. Concretely, for each step, its premises are identified by querying an LLM with the complete reasoning context preceding that step.

\subsection{Factual Error Detection}
\begin{figure*}
    \centering
    \includegraphics[width=1\textwidth]{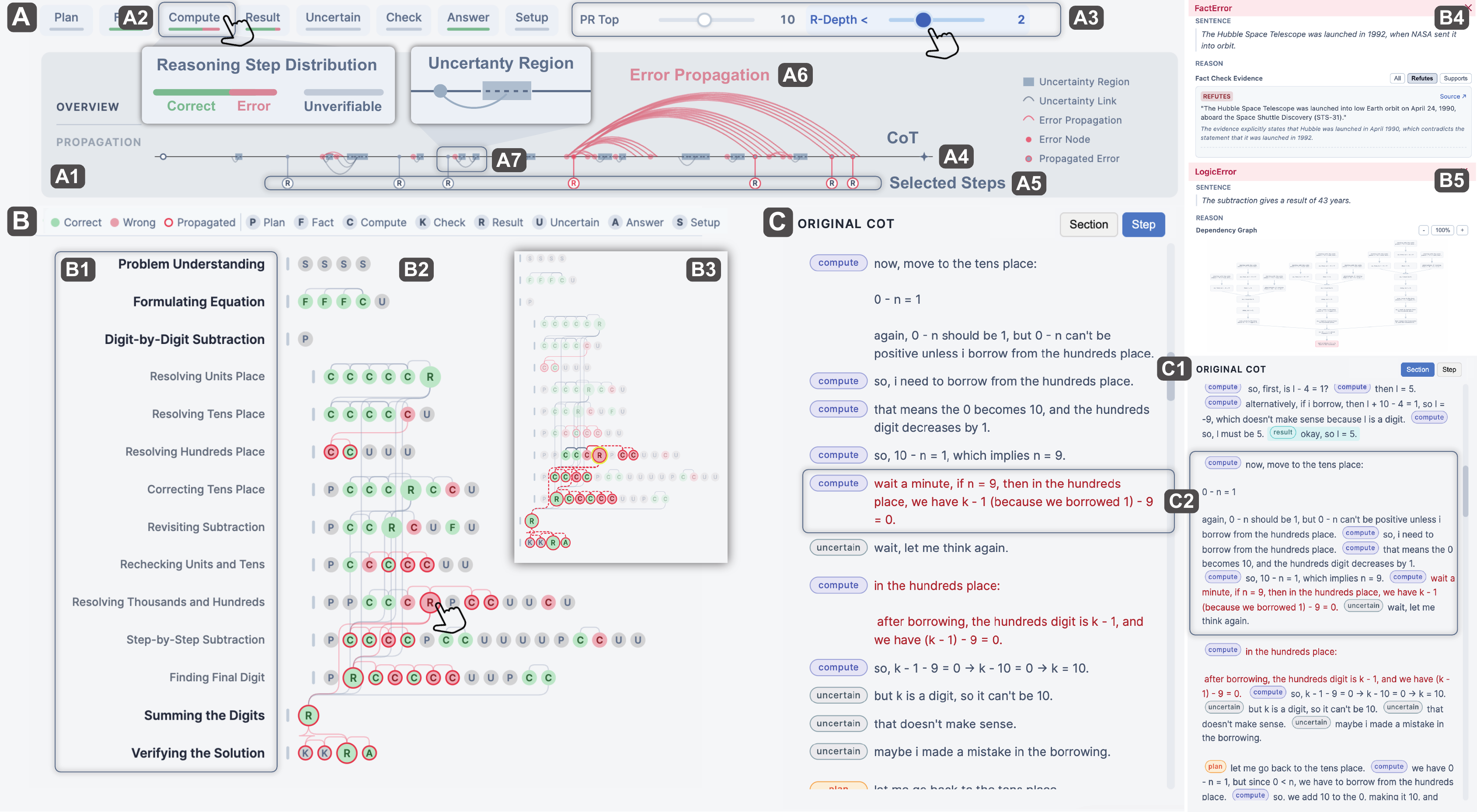}
    \caption{\name interface: (A) The \Overview displays ordinal reasoning steps (A1) along a horizontal axis (A4) and highlights uncertain regions (A7) and error propagation (A6). Users can adjust the shown steps (A5) using two filters (A2, A3). (B) The \SecView presents a hierarchical summary through textual section labels (B1) and colored step markers (B2), allowing users to click on erroneous steps to reveal their premise–conclusion relationships (B3) and the associated diagnostic evidence (B4, B5). (C) The \OriView provides the full textual CoT, organized either by individual reasoning steps or by sections for contextual inspection (C1, C2).
    }
    \vspace{-3mm}
    \label{fig:interface} 
\end{figure*}

Verifiable steps are subsequently subjected to a factual integrity analysis. This module addresses the known tendency of LRMs to generate plausible but incorrect information~\cite{ji2023survey}. The goal is to ensure that reasoning steps are grounded in verifiable external evidence (R1).\\
To achieve this, we implement retrieval-augmented verification with \textit{Loki}~\cite{li2025loki}, an open-source fact verification tool. The core methodology involves decomposing given reasoning steps into atomic queries and retrieving evidence from trusted sources via web search (using Serper API~\cite{serperapi}). From the retrieved evidence, Loki verifies the factual correctness of each step and provides ``support'', ``refute'' and ``irrelevant'' evidence about the step, facilitating the in-depth cause analysis for factual errors (R4). \csw{If the retrieved sources for a reasoning step are conflicting or absent,
we flag this step as a factual error to maximize recall, and allow users to validate it through the interactive interface (\autoref{fig:interface}\captionID{B4}).}

\subsection{Logical Error Detection}
% \wy{1. For Section 4.2 and 4.3, which step or part of our approach ensure that our approach can have a better recall of errors? Please explicitly clarify it.
% 2. After you clarify it here, then in Section 6.1, we can say the results in Table 1 satisfy our design expectation.}
Logical errors can be detected either via external symbolic solvers \cite{chen2025llmsr, pan2023logic} or by prompting LLMs\cite{xu2024faithful, mukherjee2025premise}. We adopt the former approach, using external solvers to ensure faithful, reliable, and transparent verification over formally represented knowledge. We selected Z3 \cite{de2008z3}, an SMT solver, for its rich built-in theories, which simplify formalization, and its ability to handle complex formulae involving integers, functions, and data types.
\csw{Given the user's question, a set of premise-conclusion relations (i.e., a conclusion with one or more premises), and the corresponding context (i.e., the CoT reasoning trace up to the conclusion), our goal is to verify whether the conclusion can be logically derived from the premises.} 
We first translate natural-language components into symbolic formulae using the natural-language understanding capabilities of LLMs. We design a detailed prompting strategy, drawing inspiration from prior work \cite{chen2025llmsr, pan2023logic}, that includes Z3 grammar rules and few-shot examples to guide the LLM in producing well-formed symbolic statements. Unlike prior work \cite{chen2025llmsr} that relies on iterative compilation and error correction, we employ an LLM call to convert symbolic expressions into executable Python code compatible with Z3. This approach accommodates minor symbolic variations while preserving correctness. The resulting Python program encodes the conclusion as a logical formula and the premises as constraints, and then uses the \texttt{z3-solver} library to check satisfiability. Any minor logical leap or unstated assumption triggers an unsatisfiable result in Z3. This approach aims to cover a broader range of potential logical errors for user inspection.

\section{\name Interface}
Guided by R2-R5, we propose an interactive \name interface that can be embedded into common LLM chat environments to replace the plain-text CoT, enabling users to understand the reasoning process and effectively diagnose errors.
Specifically, \name interface consists of three components (\autoref{fig:interface}): the \Overview (\autoref{fig:interface}\captionID{A}), the \SecView (\autoref{fig:interface}\captionID{B}), and the \OriView (\autoref{fig:interface}\captionID{C}). 
The \Overview visualizes step distributions (R2) and error propagation (R4) across the entire reasoning process.
The \SecView provides hierarchical structural summaries for understanding the reasoning flow (R2) and integrates premise-conclusion relations (R3) and error details (R4) for deeper diagnosis, where clicking an erroneous step reveals a pop-up window explaining its cause (R5).
The \OriView displays the textual CoT with highlighted step types and errors.
% The details of each component are elaborated in the following subsections.

\subsection{\Overview}
The \Overview provides a concise overview of the entire reasoning process using an enhanced sequence visual design (\autoref{fig:interface}\captionID{A1}) and includes two interactive filtering components (\autoref{fig:interface}\captionID{A2} and \captionID{A3}).
A horizontal line (\autoref{fig:interface}\captionID{A4}) serves as the primary axis, marking all ordinal reasoning steps across the entire reasoning process.
Small circles placed along the axis represent the individual reasoning steps, with each circle connected via a short vertical line to a labeled circular icon (\autoref{fig:interface}\captionID{A5}) that denotes its reasoning step type using one letter.
To support error diagnosis, the \Overview marks erroneous steps in red and depicts their propagation using the top red arc connections (\autoref{fig:interface}\captionID{A6}), \csw{which are always drawn forward (left-to-right) to show how earlier wrong steps affect later steps.}
Since reasoning steps of type ``\textit{uncertain}'' may indicate potential issues in the reasoning process and help reveal confidence levels in the CoT, we highlight them using gray blocks and \csw{bottom gray arcs that are always drawn backward (right-to-left)} and link them to the corresponding reasoning steps they reference, as shown in ~\autoref{fig:interface}\captionID{A7}.
Further, to maintain visual clarity and support fluid navigation, the interface provides two filtering components that allow users to control which reasoning steps are shown.
The right filtering component (\autoref{fig:interface}\captionID{A3}) allows users to surface important reasoning steps using two measures: PageRank~\cite{page1999pagerank} (importance derived from premise–conclusion dependencies among all steps) and R-Depth~\cite{ribeiro2022entailment} (importance with respect to contributing to the final answer). 
With the \Overview, users can quickly review the basic reasoning flow and determine whether errors exist and how they propagate.

\subsection{\SecView}
While the \Overview allows users to quickly reveal global patterns in step types and error propagation, helping them locate reasoning segments of interest, it does not provide sufficient semantics for analyzing specific reasoning steps.
To address this, the \SecView organizes the reasoning steps into semantically meaningful hierarchical sections and provides a more structured summary of the CoT at the section level (R2), enabling users to intuitively understand the intermediate logic within each segment.
Specifically, we take all reasoning steps as input and prompt the LLM to structure them into hierarchical sections that expose the high-level logic. During this prompting process, the LLM also generates a textual summary for each section, which describe its core reasoning intent (see Appendix~\ref{appendix:prompting_details} for detail prompt).
The \SecView displays section summaries vertically along the y-axis (\autoref{fig:interface}\captionID{B1}) and visualizes the detailed reasoning steps associated with each section to their right (\autoref{fig:interface}\captionID{B2}).
Steps within each section are visualized as circles labeled with their step type, consistent with the \Overview, and arranged horizontally according to their order. 
% The indentation of each section reflects its position within the hierarchical structure.
\csw{The indentation encodes each section's depth in the hierarchy produced by the LLM.}
A red stroke around a circle denotes that the step depends on at least one erroneous premise, indicating that even if the step itself appears correct (green fill), it is still influenced by earlier errors through its dependency chain.
The radius of each circle reflects its importance score derived from the selected measures (PageRank or R-Depth) mentioned in the \Overview, contingent on the active filtering component in the \Overview. A larger radius denotes a more influential reasoning step.
To enhance traceability (R3), we draw connecting lines between circles that share a premise–conclusion relation.
If one step serves as the premise for another, a line links their corresponding circles. 
The lines involving error propagation are marked in red.
Further, users can hover over any circle to highlight all related reasoning steps that can be traced from it through the premise–conclusion relationships (\autoref{fig:interface}\captionID{B3}).
Premise steps leading to the selected step are connected with solid lines, while subsequent steps for which the selected step serves as a premise are connected with dashed lines.
When clicking on one wrong reasoning step, a pop-up window (\autoref{fig:interface}\captionID{B4} and \captionID{B5}) will appear to show the error type (logical error or factual error) and the detailed reasons why our error detection pipeline labels it as an error (R5).
For factual errors (\autoref{fig:interface}\captionID{B4}), the pop-up window presents supporting and refuting evidence with explanatory context from our error detection pipeline. In the case of logic errors (\autoref{fig:interface}\captionID{B5}), it provides the full set of premise reasoning steps related to the erroneous step to facilitate user inspection.
Compared with the long CoT text, the \SecView reduces cognitive load in understanding the reasoning flow (R2) while preserving the detailed structural relationships among steps, thereby supporting logical traceability (R3) and facilitating analysis of error propagation (R4).

\subsection{\OriView}
To help users access the original reasoning process, the \OriView displays the entire CoT in two switchable modes: \textit{Step Mode} (\autoref{fig:interface}\captionID{C}) and \textit{Section Mode} (\autoref{fig:interface}\captionID{C1}). 
In \textit{Step Mode}, the reasoning steps are listed individually from top to bottom, each annotated with a colored label indicating its step type.
In \textit{Section Mode}, the steps are grouped by their corresponding sections, providing a natural structure that allows users to perceive the reasoning hierarchy and grasp the broader logical flow.
Wrong reasoning steps are highlighted in red.
When users select a reasoning step or section via the \Overview or \SecView, the \OriView automatically scrolls to the corresponding location and highlights it for contextual inspection (\autoref{fig:interface}\captionID{C2}).

% \subsection{Interaction}

\section{Evaluation}
This section presents a comprehensive evaluation of our system’s performance and effectiveness.

\subsection{Technical Evaluation}\label{sec-tech-evaluation}
This section details our technical evaluation setup, including dataset construction, annotation procedures, and results.

\textbf{Dataset Construction and Annotation Protocol.}
One challenge of evaluating sentence-level error detection in CoTs is the lack of datasets with fine-grained annotations.
Existing datasets, such as Deltabench \cite{he-etal-2025-large}, provide only section-level labels and omit reasoning errors that may have been self-corrected later in the CoT. \csw{To address this, we constructed a dataset of 13 CoT samples from Deltabench. Each trace was segmented into sentence-level reasoning steps, yielding 2,030 sentences in total, of which 1,171 are verifiable (28–154 per sample).
Each verifiable sentence was labeled by a subset of a 10-person annotator pool (9 PhD, 1 Master's; including one author), and disagreements were resolved through discussion and final adjudication by the author-verifier.}
% To address this, we constructed a dataset of 14 CoT samples from Deltabench, comprising 1,225 sentences (28–154 per sample).
%
%
The annotators flagged both explicit and implicit errors, including self-corrected reasoning not captured in Deltabench. \csw{The annotated dataset is publicly available on HuggingFace: \href{https://huggingface.co/datasets/CoTDiagnosis/DeltaBench_CoT_Diagnosis}{CoTDiagnosis/DeltaBench\_CoT\_Diagnosis}.}

% \subsubsection{Evaluation Setup and Performance}
\textbf{Evaluation Setup.}
We evaluated our system on the constructed dataset and compared its performance with two baselines:
\begin{itemize}
    \item \textbf{DeltaBench Prompt (Sentence-Adapted Baseline)} \cite{he-etal-2025-large} evaluates errors at the section level. For comparability, we minimally adapted its official prompt to enable sentence-level judgments while preserving the original evaluation logic.
    \item \textbf{BIG-Bench Prompt (Binary Error Classification Baseline)} \cite{tyen-etal-2024-llms} provides prompting templates for sentence-level evaluation. We applied its official prompting template to output a binary (yes/no) error decision for each sentence. 
\end{itemize}
All methods were run with GPT-5 to ensure a fair comparison. Performance was evaluated against manual ground truth using precision, recall and F1, \csw{see Appendix~\ref{appendix:error_analysis} for detailed results}.

\begin{table}[t]
\centering
\caption{\csw{A comparison of sentence-level error detection performances between \name and two baseline methods.}}
\label{tab:results_comparison}
\small
\begin{tabular}{lccc}
\toprule
Method                              & Precision. & Recall. & F1    \\
\midrule
GPT-5 (BIG-Bench Prompt)            & \textbf{0.432} & 0.658 & \textbf{0.470}  \\
GPT-5 (DeltaBench Prompt)           & 0.051 & 0.041 & 0.044 \\
\textbf{\name\ (Ours)}              & 0.306 & \textbf{0.801} & 0.386 \\
\bottomrule
\end{tabular}
\end{table}

\note{\textbf{Performance Analysis.} Table \ref{tab:results_comparison} summarizes
the sentence-level error detection performances of different methods.
\name achieves a recall of 0.801, outperforming BIG-Bench prompt (0.658) and Deltabench prompt (0.041), albeit with a lower precision (0.306) and F1 score (0.386) compared with BIG-Bench prompt (precision: 0.432, F1: 0.470).
First, the results confirm the advantage of our approach, i.e., \textit{high recall of error detection}, which is designed to be achieved in our approach as clarified in Section~\ref{sec-error-dec-pipeline}.
Prioritizing recall over precision is preferable for the error detection here, because undetected errors (false negatives) will either result in misinterpretation of the CoT reasoning process or force users to do an open-ended and time-consuming search over the original long CoTs, but misclassified steps (false positives) mainly add additional verification effort over a much smaller candidate set.
Prior work also suggests users prefer recall over precision when using an AI-powered system~\cite{kocielnik2019will}.
% \wy{pls double check if my last statement is accurate in being aligned with the original paper.}
%
%
%
% Low precision 意味着需要人工检查，但对于 baseline 也是如此;
Second, the results show that both \name and two baselines have \textit{relatively low precisions}, where the best precision is 0.432 and achieved by \textit{BIG-Bench prompt}, and our precision is 0.306. It indicates the necessity of human verification, which is addressed by the interactive visualizations in our interface.
The further improvement of our error detection pipeline is left as future work.
%
% The lower precision indicates that human verification remains necessary; importantly, this is not unique to \name, as the baseline BIG-Bench prompt (0.432) and Deltabench prompt (0.051) also requires verification due to non-trivial false positives.
% % 对于这个任务来说 Higher recall is a more important.
% Additionally, for error detection, prioritizing recall is often preferable because missed errors (false negatives) force users into open-ended search over long CoTs, whereas misclassified steps (false positives) mainly add verification effort over a much smaller candidate set. Prior work also suggests users may prefer high-recall assistants, reporting higher perceived accuracy and acceptance than high-precision variants~\cite{kocielnik2019will}.\\
% 此外，我们分析了 False Positive 的 pattern, 发现大部分都属于可以通过我们系统快速过滤的 propagated errors
%
%
%
%
%
%
% \wy{1. I do not understand why this discussion here is relevant. 2. Please check my Chinese comments here on overleaf and comments in the appendix.}
}

% \wy{This paragraph is way too messy! What is your logic flow? I am totally lost.}

\subsection{Case Study}
To further evaluate \name, we invited sixteen participants (P1-P16) to actually use \name and provide feedback through a user interview, where they were required to complete three tasks: understand and judge where the reasoning is incorrect; identify where the error occurs; and find why the error happens. 
The detailed procedure is introduced in Section~\ref{section:user_interview}. 
In this section, we present two case studies conducted by P1 and P7. 
\begin{figure*}[ht]
    \centering
    \includegraphics[width=0.95\linewidth]{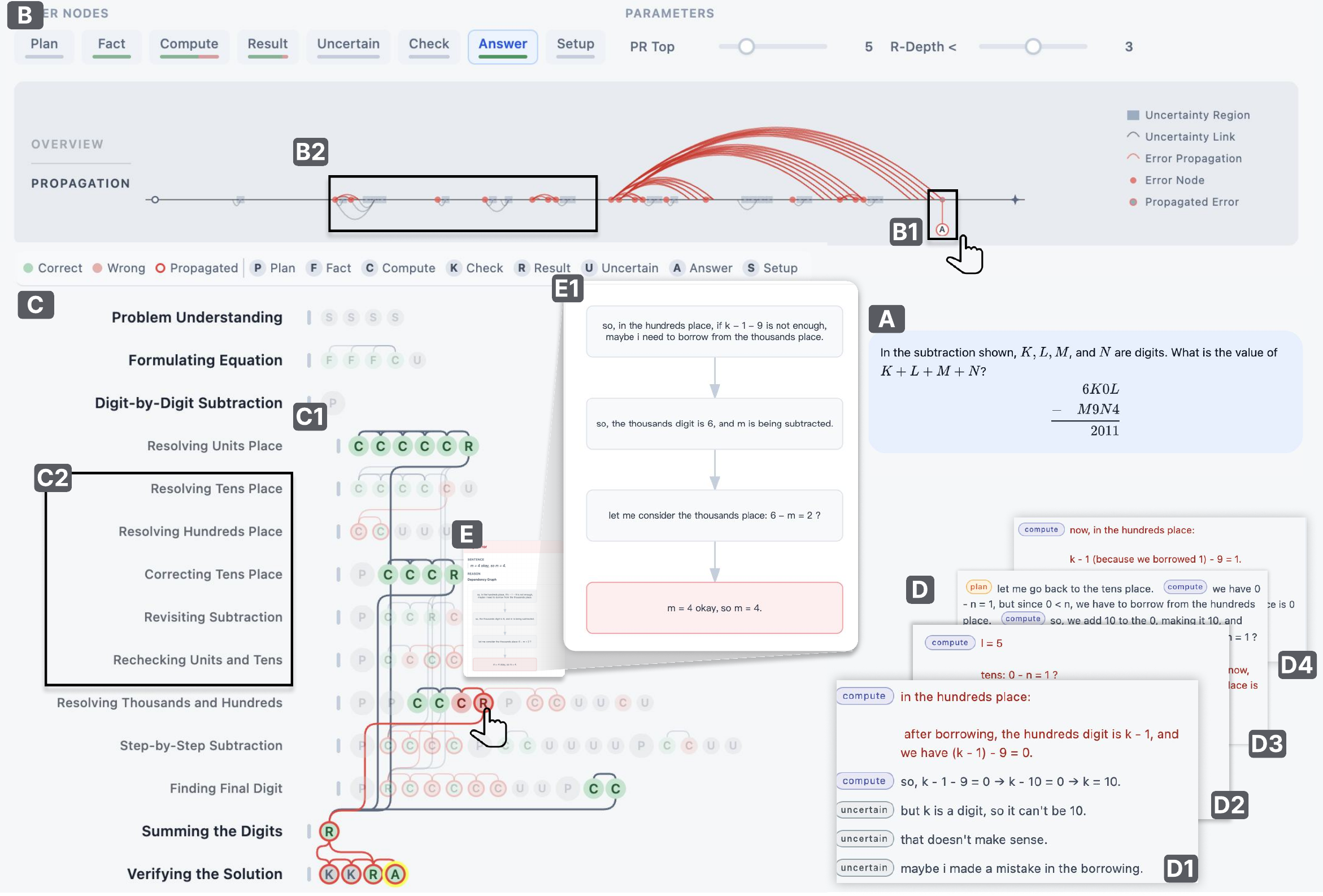}
    \caption{With \name, a user diagnoses errors and reasoning patterns in a mathematical CoT. (A) Problem statement. (B) Overview of step types and error propagation, with (B1) highlighting the ``polluted'' final answer and (B2) a retroactive reasoning pattern. (C) Structured reasoning trace, where (C1) shows the premise–conclusion chain to the answer and the red error path, and (C2) describes the self-correction phase. (D) Original CoT text, with (D1–D4) illustrating retroactive reasoning. (E) Erroneous step and its premises, with (E1) revealing a mistake where the thousands digit is not decremented.}
    \label{fig:case1}
\end{figure*}

\subsubsection{Case 1: Diagnose Error Cause and Reasoning Patterns}
\label{section:case}
P7 is an undergraduate student who majors in mathematical science but has limited knowledge about AI. \csw{She represents a representative type of our target users: 
% domain-knowledgeable but without AI expertise.
domain experts without expertise in AI.
} 
% \wy{Pls clarify the target users of \name in the introduction section.}
During the user interview, she was provided with a mathematical calculation question (\autoref{fig:case1}\captionID{A}) from DeltaBench~\cite{he-etal-2025-large}. The CoT reasoning trace for this question contains 97 reasoning steps.

\textbf{Identifying errors and tracing their causes.}
P7 began by using the interactive filtering components to examine the distribution of reasoning steps. She immediately observed that ``Compute'' steps exhibited a disproportionately high frequency of errors compared to other step types. She noted that this distribution was consistent with the computational nature of the task.
After several exploratory checks, P7 narrowed her focus to the ``Plan'', ``Results'', and ``Answer'' steps, which she considered most informative for understanding and assessing the reasoning trace. Following the tasks, she applied the ``Answer'' filter (\autoref{fig:case1}\captionID{B}). This action revealed the final answer step(\autoref{fig:case1}\captionID{B1}). The red stroke of this step indicated that it was ``polluted'' by errors propagated from earlier steps. To identify where the ``polluting'' error occurs, P7 clicked the answer node. In response, the \SecView highlighted the complete logical dependency chain (\autoref{fig:case1}\captionID{C1}), visualizing the lineage of the error. By tracing the red path upstream, she isolated the originating error node.
A detailed inspection of its premises (\autoref{fig:case1}\captionID{E}, \captionID{E1}) against the problem statement revealed the specific cognitive failure: while the   text explicitly stated the need to ``borrow one from the thousands,'' the model failed to decrement the digit in the actual calculation.

\textbf{Retroactive reasoning patterns.}
During her exploratory checks in the \Overview, P7 observed a characteristic pattern: she noticed that certain steps were marked as errors but did not produce long propagation arcs (\autoref{fig:case1}\captionID{B2}). Instead, they were followed by one or more uncertainty steps, each connected back to the erroneous node through an uncertainty link. This pattern suggested that the model had detected its own errors and repeatedly attempted local self-corrections. 
Curious about what exactly the model was trying to correct, P7 then navigated into these regions in the \SecView. By reading the left-hand reasoning text, she concluded that the model was following a digit-by-digit computation strategy (\autoref{fig:case1}\captionID{C2}), repeatedly recomputing values at the tens, hundreds, and thousands places. As she clicked through individual sections and aligned them with the original reasoning text, P7 discovered that an earlier borrow-related mistake caused the model to generate impossible intermediate values (e.g., $k = 10$) (\autoref{fig:case1}\captionID{D1}). Each time such an impossibility appeared, she observed the model going back to tweak previous digits. \csw{This finding shows that the model actually made several successful attempts at self-correction}.

\subsubsection{Case 2: Diagnose Illusory Truth and Logical Gaps}
\label{section:case2}
\begin{figure}[ht]
    \centering
    \includegraphics[width=\linewidth]{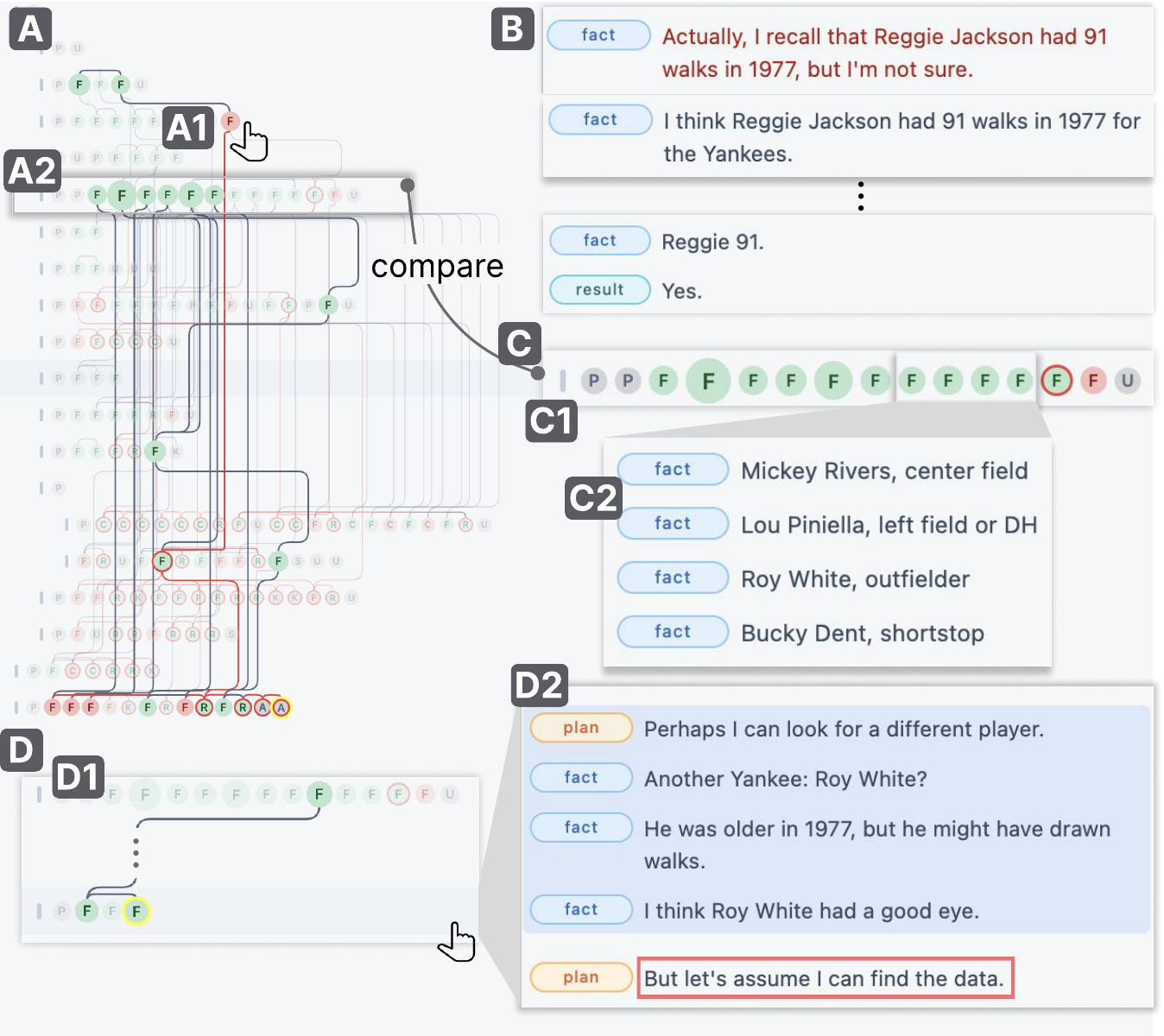}
    % \caption{Case 2.}
    \caption{With \name, a user diagnoses illusory truth and logical gaps. (A) Structured reasoning trace where only a sparse subset of steps contributes to the final answer; (A1) marks the key factual error and (A2) a disconnected high-importance section. (B) Shows the model’s shift from tentative to definitive statements, illustrating the illusory truth effect. (C) Highlights unused player statistics. (D) Shows a terminated reasoning branch for ``Roy White,'' indicating a logical gap.}

    % \caption{With \name, a user diagnoses illusory truth and logical gaps. (A) shows the structured reasoning trace revealing that only a sparse subset of steps contributes to the final answer, where (A1) highlights the key factual error influencing the final answer and (A2) surfaces a disconnected high-importance section. (B) shows the model's shift from tentative to definitive. (C) reveals certain player's statics not used. (D) reveals a terminated reasoning branch for ``Roy White,'' indicating a logical gap.}
    \label{fig:case2}
\end{figure}
% P1 is an PhD student majoring in machine learning. The second case is about a multi-hop query from the GAIA asking for ``How many at bats did the Yankee with the most walks in the 1977 regular season have that same season?''. Having the diagnosis experience as Case 1, P1 quickly identify that the factual error at (\autoref{fig:case2}\captionID{A1}) and several factual error at the last section is the cause. 
P1 is a PhD student specializing in machine learning. The second case concerns a multi-hop query from the GAIA benchmark asking, \textit{``How many at bats did the Yankee with the most walks in the 1977 regular season have that same season?''}, \csw{and the trace is generated by Deepseek-R1. 
P1 represents
another type of our representative target users: LRM practitioners without sufficient domain knowledge for the question.
}
Drawing on the diagnosis experience from Case 1, P1 quickly identifies that the factual error at (\autoref{fig:case2}\captionID{A1}) and several factual errors in the final section are the primary causes.
% \textbf{"三人成虎"现象} 
% sparse info, 观察到很少的节点真正在获取结果时使用，从起点开始看发现模型在开始表现出的不确定性突变为确认，很奇怪。因此用户观察这个数据在后续使用过程，同不断点击使用同样数据信息，观察到虽然第一次出现的时候模型表明了uncertainty，但随着后续的不断出现这个值，虽然没有任何让模型能够确认的信息源，但模型直接把之前不确定的句子判断。

\textbf{Illusory truth effect.} By tracing the lineage of the error in \autoref{fig:case2}\captionID{A}, P1 initially noticed that the final answer relied on a surprisingly sparse set of steps. Upon inspecting the text content of the starting node (\autoref{fig:case2}\captionID{B}), he noted that the model explicitly qualified the statistics with uncertainty, using phrases like ``but I'm not sure'' or ``I recall...''. However, as he traced the following steps that use this step as a premise, he noticed a dramatic shift from tentative to definitive. He noted that as the information was accessed and restated, despite the absence of any new external evidence to verify the claim, the model treated its own previous uncertain generation as a confirmed fact, creating a self-reinforcing loop of error.

% \textbf{逻辑跳跃}
% 通常以result或者uncertainty结尾，发现一个section中的句子在进行P和F，C后就结束, 并且F在后续没有被使用，
\textbf{Logical Gaps.}
To investigate the correct answer, P1 filtered nodes based on PageRank. He identified a high-ranking section listing potential players that was disconnected from the final answer (\autoref{fig:case2}\captionID{A2},\captionID{C1}). Tracing the specific branch for each disconnected step, he observed the branch for ``Roy White'' terminated abruptly without resolving to a result or uncertainty state (\autoref{fig:case2}\captionID{D1}). 
Checking the actual reasoning text (\autoref{fig:case2}\captionID{D2}), P1 saw that the model had gathered supporting facts, but abruptly abandoned this reasoning thread without performing the necessary comparison or verification step.

\subsection{User Interview}
\label{section:user_interview}

We conducted semi-structured user interviews with 16 participants to evaluate the effectiveness and usability of \name.

\subsubsection{Participants}
Sixteen participants (P1-P16) were recruited from a local university. Ages ranged from 18 to 31 years. \csw{As our target users encompass a broad range of LRM users, who need to inspect CoT traces, we purposefully selected a diverse cohort to capture the real-world variance in terms of their backgrounds in the application domains and their expertise in AI.} The sample comprised 6 male and 10 female participants from diverse disciplines, including Finance (3), Electronic Engineering (3),  Human–Computer Interaction (3), Mathematical Science (2), Business (1), Machine Learning (1), Economics (1), Biochemistry and Molecular Biology (1), and Ocean Engineering (1). All participants had at least one year of experience using LRMs and reported high familiarity with LRMs (12/16 indicated nearly daily use). The interviews were conducted online via Zoom, and \name was accessed on participants’ own laptops or desktop computers. \csw{The user study materials are available at \href{https://csw0109.github.io/reasondiag-demo/}{reasondiag-demo.github.io}}.

\begin{figure*}[ht]
    \centering
    \includegraphics[width=0.95\linewidth]{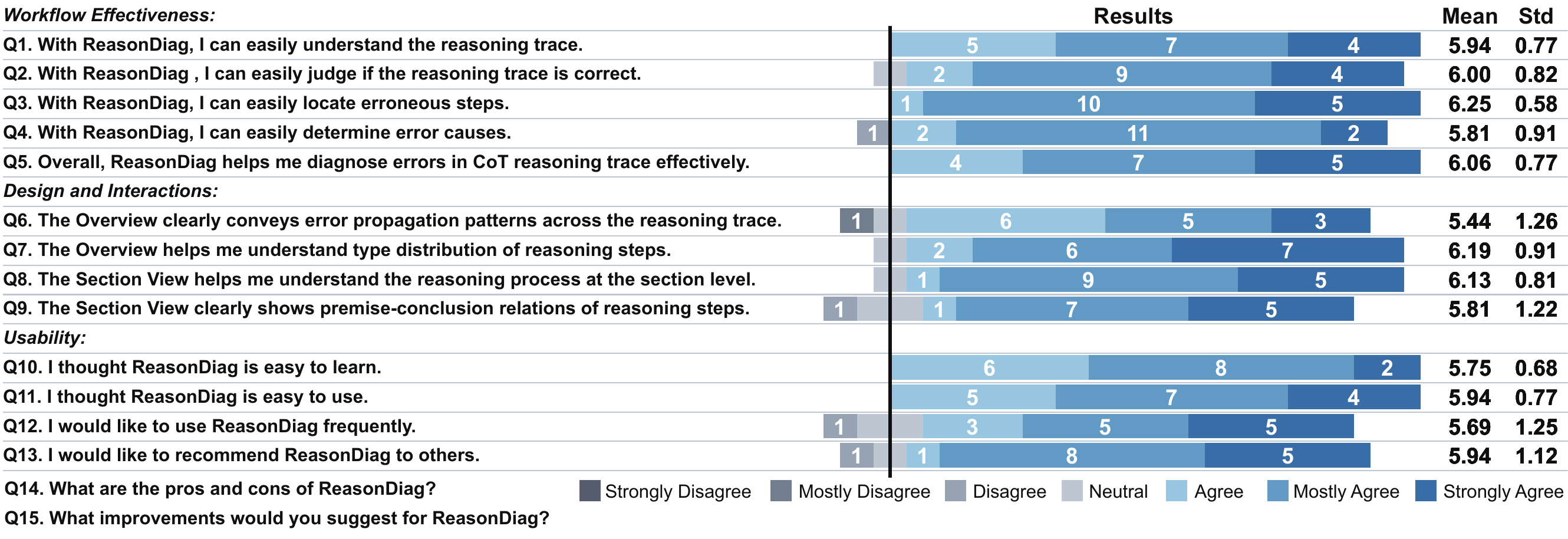}
    \caption{The user interview questionnaire results. Q1-Q13 are closed-ended and rated on a 7-point Likert scale. Q14 and Q15 are open-ended questions to collect participants’ feedback. The detailed scores of Q1-Q10 are shown in a stacked bar chart.}
    \label{fig:interview}
\end{figure*}
\subsubsection{Procedures}
Each study session lasted approximately 50 minutes and was organized into four distinct stages. First, we introduced the study background and defined the three core diagnostic tasks that participants would perform: (1) \textit{Comprehension and Judgment}, understanding the reasoning trace and determining the correctness of the solution; (2) \textit{Error Localization}, locating where errors occurred; and (3) \textit{Root Cause Analysis}, explaining why these errors occurred. Second, we conducted a guided walkthrough of the main functions provided by \name. Third, participants were given a period of free exploration to familiarize themselves with the interface and interaction mechanisms. Finally, participants were given two diagnostic cases to evaluate using the system. These cases represented different reasoning modalities:
\textbf{Case 1:} A mathematical reasoning sample from the DeltaBench dataset (\autoref{fig:case1}\captionID{A}).
\textbf{Case 2:} A multi-hop query from the GAIA benchmark~\cite{mialon2023gaia} whose reasoning trace was generated by DeepSeek-R1. 
Participants are encouraged to think aloud throughout the process. Upon completing the tasks, participants completed a 7-point Likert-scale questionnaire and a semi-structured interview to provide qualitative feedback. Each participant received \$10 compensation for their time.

\subsubsection{Results}
~\autoref{fig:interview} shows user feedback on \name along four aspects:

\textbf{Workflow Effectiveness:} Participants rated the workflow highly, with all items scoring above 5.8. The tool is good at locating erroneous steps (Q3, $M=6.25$ ), proven effective for pinpointing failures in long reasoning traces. Effectiveness in diagnosing errors (Q5, $M=6.06$) and judging correctness (Q2, $M=6.00$) was also confirmed. However, qualitative feedback highlighted that the inherent complexity of logical reasoning remains a challenge for some users. One participant (P10), who gave a lower rating for determining error causes, noted that identifying the root cause (Q4) still demands a ``fair amount of reasoning.'' P10 explained that while the visualization brings the error into view, bridging the semantic gap is not always immediate.

\textbf{Design and Interaction:} 
The visual design components received generally high praise, particularly regarding the distribution analysis. Participants agreed that the \Overview effectively conveys the type distribution of reasoning steps (Q7, $M=6.19, SD=0.91$). The \SecView was also highly valued for helping users understand the reasoning process at a local level (Q8, $M=6.13, SD=0.81$).
However, we observed slightly higher variance in participants' feedback regarding the clarity of error propagation patterns in the \Overview (Q6, $M=5.44, SD=1.26$) and the premise-conclusion relations in the \SecView (Q9, $M=5.81, SD=1.22$). From the semi-structured interview, we gather that while most users find the visualizations effective for high-level tasks, certain dense logical structures may present readability challenges, which we discuss in the limitations.

\textbf{Usability:} 
Participants found \name user-friendly, rating it highly for being easy to use (Q11, $M=5.94$) and easy to learn (Q10, $M=5.75$). This accessibility translated into a willingness to recommend the tool (Q13, $M=5.94$).
Notably, the intention to use frequently (Q12, $M=5.69$) showed higher variance ($SD=1.25$). Qualitative feedback suggests this reflects the specialized nature of the task: debugging CoT is an occasional necessity rather than a daily routine, yet users find the tool valuable when needed.

\textbf{Limitations and improvements:}
Although participants generally found \name effective for diagnosing errors in long reasoning traces, several limitations and improvements were identified. A common concern was visual complexity caused by dense links between steps. 
When a reasoning step influences numerous subsequent steps, the resulting horizontal and diagonal edges accumulate rapidly, obscuring the underlying structure. As P15 noted, \textit{``once there are many cross-line relations, it becomes impossible to tell which node connects to which''}. 
Beyond visualization, participants expressed a desire for the system to move from passive diagnosis toward active support. Several noted that the current design \textit{``only marks it but does not resolve it,''} and called for more fine-grained distinctions between logic error types (e.g., ``evidence missing'' and ``evidence contradicting''). As future directions, participants suggested providing direct modification to the reasoning trace, like correcting erroneous CoT reasoning steps or simplifying repetitive reasoning.

\section{Discussion and Future Work}
% Here we reflect on the limitations of our approach, potential for generalization, and directions for future improvement.
Despite the effectiveness of \name demonstrated above, we further discuss the possible limitations and future work.

\textbf{Scalability and Granularity Trade-offs}.
The visualization of multi-step reasoning traces presents inherent challenges regarding computational cost and information density. Complex tasks often necessitate extensive sequences of intermediate steps. For instance, reasoning traces in the DeltaBench~\cite{he-etal-2025-large} dataset frequently exceed 100 steps. This high granularity poses a dual challenge: computationally, it increases system latency; visually, it clutters the \Overview.
We observed that fixed, step-level granularity often treats simple operations as individual nodes, inflating graph size without adding semantic depth.
To address this, our future work will explore adaptive granularity algorithms that aggregate simple operations into composite nodes while preserving detail for complex branches, improving scalability, and reducing low-level noise.

\note{
\textbf{Error Detection Precision.}
As discussed in Section~\ref{sec-tech-evaluation}, \name’s error-detection pipeline achieves high recall over baseline methods for finding logical and factual errors in CoT reasoning traces, but its precision is lower. Our error analysis (Appendix~\ref{appendix:error_analysis}) shows that many false positives come from error propagation (44.48\%): when a few early core steps are misclassified, many later steps are flagged as errors as well. In practice, \name reduces this overhead by helping users trace a burst of flagged steps back to the original core steps for quick checking. In future work, we can use agreement across multiple verification methods (e.g., our pipeline and prompt-based verification) to reduce these false positives while preserving high recall. In addition, it is worth exploring how more advanced error detection approaches can be incorporated into \name, further enhancing its usability.
}

\textbf{Cognitive Load and the Semantic Gap}.
While our visualization significantly reduces the effort for error localization, semantic verification still imposes substantial cognitive load. Our user study reveals a dichotomy in debugging efficiency based on error complexity.
For explicit factual errors, such as using incorrect information, with the provided fact-check evidence, participants could diagnose the issue immediately. However, subtle logical errors often require more time and deeper reasoning, sometimes with domain knowledge.
 This ``semantic gap'' indicates that locating an anomalous node is easier than explaining \emph{why} it is wrong: users must reconstruct the logic and cross-check the problem context. Echoing P10’s feedback, future work should add AI-assisted explanations for logic errors and \csw{support interactive debugging (e.g., editing erroneous steps) to help users locate and also fix reasoning errors. }

\textbf{Generalizability across Domains}.

Our evaluation targets domains with verifiable ground truth (math, logic, information retrieval). While our visual design principles are intended to be domain-agnostic, their effectiveness in less structured settings (e.g., creative writing or legal argumentation) remains unvalidated. In math, correctness is often binary and supports clear error marking (e.g., red nodes); in subjective domains, “errors” are ambiguous and may require softer encodings of uncertainty or plausibility. Future work will test how our node-link metaphors and interactions transfer to open-ended contexts, potentially adding encodings for semantic drift or tonal inconsistency.

\section{Conclusion}

In this work, we presented \name, an interactive system for diagnosing CoT reasoning traces, together with an automatic pipeline that combines external fact-checking and symbolic formal logic to detect step-level factual and logical errors. We grounded our design with a formative study with 9 participants. \name integrates an arc diagram to reveal reasoning step distributions and error-propagation patterns, and a hierarchical node-link diagram to show high-level reasoning flows and premise dependencies. Our 
extensive evaluations have
demonstrated that \name helps users understand long CoT reasoning traces, identify erroneous steps, and analyze their root causes, offering a step toward more trustworthy LLM reasoning.

\section*{Acknowledgements}
We thank the anonymous reviewers for their constructive feedback and suggestions. Generative AI tools were used in a limited manner solely for grammar correction and language polishing. This project was supported by the Ministry of Education, Singapore, under its Academic Research Fund Tier 2 (Proposal ID: T2EP202220049), and by the NTU Start Up Grant awarded to Yong Wang.
Any opinions, findings, conclusions, or recommendations expressed in this material are those of the author(s) and do not necessarily reflect the views of the Ministry of Education, Singapore.

\bibliographystyle{eg-alpha-doi} 
\bibliography{egbibsample}

@techreport{chatterji2025people,
  title={How people use chatgpt},
  author={Chatterji, Aaron and Cunningham, Thomas and Deming, David J and Hitzig, Zoe and Ong, Christopher and Shan, Carl Yan and Wadman, Kevin},
  year={2025},
  institution={National Bureau of Economic Research}
}

@article{wei2022chain,
  title={Chain-of-thought prompting elicits reasoning in large language models},
  author={Wei, Jason and Wang, Xuezhi and Schuurmans, Dale and Bosma, Maarten and Xia, Fei and Chi, Ed and Le, Quoc V and Zhou, Denny and others},
  journal={Advances in neural information processing systems},
  volume={35},
  pages={24824--24837},
  year={2022}
}

@article{barez2025chain,
  title={Chain-of-thought is not explainability},
  author={Barez, Fazl and Wu, Tung-Yu and Arcuschin, Iv{\'a}n and Lan, Michael and Wang, Vincent and Siegel, Noah and Collignon, Nicolas and Neo, Clement and Lee, Isabelle and Paren, Alasdair and others},
  journal={Preprint, alphaXiv},
  pages={v1},
  year={2025}
}

@inproceedings{chen2025llmsr,
  title={LLMSR@ XLLM25: SWRV: Empowering Self-Verification of Small Language Models through Step-wise Reasoning and Verification},
  author={Chen, Danchun},
  booktitle={Proceedings of the 1st Joint Workshop on Large Language Models and Structure Modeling (XLLM 2025)},
  pages={322--335},
  year={2025}
}

@inproceedings{pan2023logic,
  title={Logic-lm: Empowering large language models with symbolic solvers for faithful logical reasoning},
  author={Pan, Liangming and Albalak, Alon and Wang, Xinyi and Wang, William},
  booktitle={Findings of the Association for Computational Linguistics: EMNLP 2023},
  pages={3806--3824},
  year={2023}
}

@inproceedings{mukherjee2025premise,
title={Premise-Augmented Reasoning Chains Improve Error Identification in Math reasoning with {LLM}s},
author={Sagnik Mukherjee and Abhinav Chinta and Takyoung Kim and Tarun Anoop Sharma and Dilek Hakkani Tur},
booktitle={Forty-second International Conference on Machine Learning},
year={2025},
url={https://openreview.net/forum?id=4tYckHNVXV}
}

@inproceedings{de2008z3,
  title={Z3: An efficient SMT solver},
  author={De Moura, Leonardo and Bj{\o}rner, Nikolaj},
  booktitle={International conference on Tools and Algorithms for the Construction and Analysis of Systems},
  pages={337--340},
  year={2008},
  organization={Springer}
}

@inproceedings{jiang2023graphologue,
  title={Graphologue: Exploring large language model responses with interactive diagrams},
  author={Jiang, Peiling and Rayan, Jude and Dow, Steven P and Xia, Haijun},
  booktitle={Proceedings of the 36th annual ACM symposium on user interface software and technology},
  pages={1--20},
  year={2023}
}

@article{wang2023commonsensevis,
  title={Commonsensevis: Visualizing and understanding commonsense reasoning capabilities of natural language models},
  author={Wang, Xingbo and Huang, Renfei and Jin, Zhihua and Fang, Tianqing and Qu, Huamin},
  journal={IEEE Transactions on Visualization and Computer Graphics},
  volume={30},
  number={1},
  pages={273--283},
  year={2023},
  publisher={IEEE}
}

@article{zhou2025landscape,
  title={Landscape of thoughts: Visualizing the reasoning process of large language models},
  author={Zhou, Zhanke and Zhu, Zhaocheng and Li, Xuan and Galkin, Mikhail and Feng, Xiao and Koyejo, Sanmi and Tang, Jian and Han, Bo},
  journal={arXiv preprint arXiv:2503.22165},
  year={2025}
}

@article{pang2025interactive,
  title={Interactive Reasoning: Visualizing and Controlling Chain-of-Thought Reasoning in Large Language Models},
  author={Pang, Rock Yuren and Feng, KJ and Feng, Shangbin and Li, Chu and Shi, Weijia and Tsvetkov, Yulia and Heer, Jeffrey and Reinecke, Katharina},
  journal={arXiv preprint arXiv:2506.23678},
  year={2025}
}

@article{korbak2025chain,
  title={Chain of thought monitorability: A new and fragile opportunity for ai safety},
  author={Korbak, Tomek and Balesni, Mikita and Barnes, Elizabeth and Bengio, Yoshua and Benton, Joe and Bloom, Joseph and Chen, Mark and Cooney, Alan and Dafoe, Allan and Dragan, Anca and others},
  journal={arXiv preprint arXiv:2507.11473},
  year={2025}
}

@inproceedings{lightman2023let,
  title={Let's verify step by step},
  author={Lightman, Hunter and Kosaraju, Vineet and Burda, Yuri and Edwards, Harrison and Baker, Bowen and Lee, Teddy and Leike, Jan and Schulman, John and Sutskever, Ilya and Cobbe, Karl},
  booktitle={The Twelfth International Conference on Learning Representations},
  year={2023}
}

@inproceedings{wangself,
  title={Self-Consistency Improves Chain of Thought Reasoning in Language Models},
  author={Wang, Xuezhi and Wei, Jason and Schuurmans, Dale and Le, Quoc V and Chi, Ed H and Narang, Sharan and Chowdhery, Aakanksha and Zhou, Denny},
  booktitle={The Eleventh International Conference on Learning Representations}
}

@inproceedings{xu2024faithful,
  title={Faithful Logical Reasoning via Symbolic Chain-of-Thought},
  author={Xu, Jundong and Fei, Hao and Pan, Liangming and Liu, Qian and Lee, Mong-Li and Hsu, Wynne},
  booktitle={Proceedings of the 62nd Annual Meeting of the Association for Computational Linguistics (Volume 1: Long Papers)},
  pages={13326--13365},
  year={2024}
}

@inproceedings{li-etal-2025-reasongraph,
    title = "{R}eason{G}raph: Visualization of Reasoning Methods and Extended Inference Paths",
    author = "Li, Zongqian  and
      Shareghi, Ehsan  and
      Collier, Nigel",
    editor = "Mishra, Pushkar  and
      Muresan, Smaranda  and
      Yu, Tao",
    booktitle = "Proceedings of the 63rd Annual Meeting of the Association for Computational Linguistics (Volume 3: System Demonstrations)",
    month = jul,
    year = "2025",
    address = "Vienna, Austria",
    publisher = "Association for Computational Linguistics",
    doi = "10.18653/v1/2025.acl-demo.14",
    pages = "140--147",
    ISBN = "979-8-89176-253-4",
}

@inproceedings{tyen-etal-2024-llms,
    title = "{LLM}s cannot find reasoning errors, but can correct them given the error location",
    author = "Tyen, Gladys  and
      Mansoor, Hassan  and
      Carbune, Victor  and
      Chen, Peter  and
      Mak, Tony",
    editor = "Ku, Lun-Wei  and
      Martins, Andre  and
      Srikumar, Vivek",
    booktitle = "Findings of the Association for Computational Linguistics: ACL 2024",
    month = aug,
    year = "2024",
    address = "Bangkok, Thailand",
    publisher = "Association for Computational Linguistics",
    doi = "10.18653/v1/2024.findings-acl.826",
    pages = "13894--13908",
}

@misc{chatgpt2023,
    author = {OpenAI},
    title = {ChatGPT: Language Model},
    year = {2023},
    url = {https://chatgpt.com/},
    howpublished = {Accessed: 2023-10-01}
}

@misc{serperapi,
    author = {Serper API},
    title = {Serper API},
    url = {https://serper.dev/},
}

@article{bogdan2025thought,
  title={Thought Anchors: Which LLM Reasoning Steps Matter?},
  author={Bogdan, Paul C and Macar, Uzay and Nanda, Neel and Conmy, Arthur},
  journal={arXiv preprint arXiv:2506.19143},
  year={2025}
}

@inproceedings{venhoffunderstanding,
  title={Understanding Reasoning in Thinking Language Models via Steering Vectors},
  author={Venhoff, Constantin and Arcuschin, Iv{\'a}n and Torr, Philip and Conmy, Arthur and Nanda, Neel},
  booktitle={Workshop on Reasoning and Planning for Large Language Models}
}

@inproceedings{majer2024claim,
    title = "Claim Check-Worthiness Detection: How Well do {LLM}s Grasp Annotation Guidelines?",
    author = "Majer, Laura  and
      {\v{S}}najder, Jan",
    editor = "Schlichtkrull, Michael  and
      Chen, Yulong  and
      Whitehouse, Chenxi  and
      Deng, Zhenyun  and
      Akhtar, Mubashara  and
      Aly, Rami  and
      Guo, Zhijiang  and
      Christodoulopoulos, Christos  and
      Cocarascu, Oana  and
      Mittal, Arpit  and
      Thorne, James  and
      Vlachos, Andreas",
    booktitle = "Proceedings of the Seventh Fact Extraction and VERification Workshop (FEVER)",
    month = nov,
    year = "2024",
    address = "Miami, Florida, USA",
    publisher = "Association for Computational Linguistics",
    doi = "10.18653/v1/2024.fever-1.27",
    pages = "245--263"
}

@inproceedings{li2025loki,
    title = "Loki: An Open-Source Tool for Fact Verification",
    author = "Li, Haonan  and
      Han, Xudong  and
      Wang, Hao  and
      Wang, Yuxia  and
      Wang, Minghan  and
      Xing, Rui  and
      Geng, Yilin  and
      Zhai, Zenan  and
      Nakov, Preslav  and
      Baldwin, Timothy",
    editor = "Rambow, Owen  and
      Wanner, Leo  and
      Apidianaki, Marianna  and
      Al-Khalifa, Hend  and
      Eugenio, Barbara Di  and
      Schockaert, Steven  and
      Mather, Brodie  and
      Dras, Mark",
    booktitle = "Proceedings of the 31st International Conference on Computational Linguistics: System Demonstrations",
    month = jan,
    year = "2025",
    address = "Abu Dhabi, UAE",
    publisher = "Association for Computational Linguistics",
    pages = "28--36",
}

@article{ji2023survey,
author = {Ji, Ziwei and Lee, Nayeon and Frieske, Rita and Yu, Tiezheng and Su, Dan and Xu, Yan and Ishii, Etsuko and Bang, Ye Jin and Madotto, Andrea and Fung, Pascale},
title = {Survey of Hallucination in Natural Language Generation},
year = {2023},
issue_date = {December 2023},
publisher = {Association for Computing Machinery},
address = {New York, NY, USA},
volume = {55},
number = {12},
issn = {0360-0300},
url = {https://doi.org/10.1145/3571730},
doi = {10.1145/3571730},
journal = {ACM Comput. Surv.},
month = mar,
articleno = {248},
numpages = {38}
}

@misc{baker2025monitor,
      title={Monitoring Reasoning Models for Misbehavior and the Risks of Promoting Obfuscation}, 
      author={Bowen Baker and Joost Huizinga and Leo Gao and Zehao Dou and Melody Y. Guan and Aleksander Madry and Wojciech Zaremba and Jakub Pachocki and David Farhi},
      year={2025},
      eprint={2503.11926},
      archivePrefix={arXiv},
      primaryClass={cs.AI},
      url={https://arxiv.org/abs/2503.11926}, 
}

@misc{qwen,
      title={Qwen3 Technical Report}, 
      author={An Yang and Anfeng Li and Baosong Yang and Beichen Zhang and Binyuan Hui and Bo Zheng and Bowen Yu and Chang Gao and Chengen Huang and Chenxu Lv and Chujie Zheng and Dayiheng Liu and Fan Zhou and Fei Huang and Feng Hu and Hao Ge and Haoran Wei and Huan Lin and Jialong Tang and Jian Yang and Jianhong Tu and Jianwei Zhang and Jianxin Yang and Jiaxi Yang and Jing Zhou and Jingren Zhou and Junyang Lin and Kai Dang and Keqin Bao and Kexin Yang and Le Yu and Lianghao Deng and Mei Li and Mingfeng Xue and Mingze Li and Pei Zhang and Peng Wang and Qin Zhu and Rui Men and Ruize Gao and Shixuan Liu and Shuang Luo and Tianhao Li and Tianyi Tang and Wenbiao Yin and Xingzhang Ren and Xinyu Wang and Xinyu Zhang and Xuancheng Ren and Yang Fan and Yang Su and Yichang Zhang and Yinger Zhang and Yu Wan and Yuqiong Liu and Zekun Wang and Zeyu Cui and Zhenru Zhang and Zhipeng Zhou and Zihan Qiu},
      year={2025},
      eprint={2505.09388},
      archivePrefix={arXiv},
      primaryClass={cs.CL},
      url={https://arxiv.org/abs/2505.09388}, 
}

@inproceedings{he-etal-2025-large,
    title = "Can Large Language Models Detect Errors in Long Chain-of-Thought Reasoning?",
    author = "He, Yancheng  and
      Li, Shilong  and
      Liu, Jiaheng  and
      Wang, Weixun  and
      Bu, Xingyuan  and
      Zhang, Ge  and
      Peng, Z.y.  and
      Zhang, Zhaoxiang  and
      Zheng, Zhicheng  and
      Su, Wenbo  and
      Zheng, Bo",
    editor = "Che, Wanxiang  and
      Nabende, Joyce  and
      Shutova, Ekaterina  and
      Pilehvar, Mohammad Taher",
    booktitle = "Proceedings of the 63rd Annual Meeting of the Association for Computational Linguistics (Volume 1: Long Papers)",
    month = jul,
    year = "2025",
    address = "Vienna, Austria",
    publisher = "Association for Computational Linguistics",
    doi = "10.18653/v1/2025.acl-long.905",
    pages = "18468--18489",
    ISBN = "979-8-89176-251-0",
}

@article{guo2025deepseek,
  title={Deepseek-r1: Incentivizing reasoning capability in llms via reinforcement learning},
  author={Guo, Daya and Yang, Dejian and Zhang, Haowei and Song, Junxiao and Zhang, Ruoyu and Xu, Runxin and Zhu, Qihao and Ma, Shirong and Wang, Peiyi and Bi, Xiao and others},
  journal={arXiv preprint arXiv:2501.12948},
  year={2025}
}

@article{team2023gemini,
  title={Gemini: a family of highly capable multimodal models},
  author={Team, Gemini and Anil, Rohan and Borgeaud, Sebastian and Alayrac, Jean-Baptiste and Yu, Jiahui and Soricut, Radu and Schalkwyk, Johan and Dai, Andrew M and Hauth, Anja and Millican, Katie and others},
  journal={arXiv preprint arXiv:2312.11805},
  year={2023}
}

@inproceedings{mialon2023gaia,
  title={Gaia: a benchmark for general ai assistants},
  author={Mialon, Gr{\'e}goire and Fourrier, Cl{\'e}mentine and Wolf, Thomas and LeCun, Yann and Scialom, Thomas},
  booktitle={The Twelfth International Conference on Learning Representations}
}

@techreport{page1999pagerank,
  title={The PageRank citation ranking: Bringing order to the web.},
  author={Page, Lawrence and Brin, Sergey and Motwani, Rajeev and Winograd, Terry},
  year={1999},
  institution={Stanford infolab}
}

@inproceedings{ribeiro2022entailment,
    title = "Entailment Tree Explanations via Iterative Retrieval-Generation Reasoner",
    author = "Neves Ribeiro, Danilo  and
      Wang, Shen  and
      Ma, Xiaofei  and
      Dong, Rui  and
      Wei, Xiaokai  and
      Zhu, Henghui  and
      Chen, Xinchi  and
      Xu, Peng  and
      Huang, Zhiheng  and
      Arnold, Andrew  and
      Roth, Dan",
    editor = "Carpuat, Marine  and
      de Marneffe, Marie-Catherine  and
      Meza Ruiz, Ivan Vladimir",
    booktitle = "Findings of the Association for Computational Linguistics: NAACL 2022",
    month = jul,
    year = "2022",
    address = "Seattle, United States",
    publisher = "Association for Computational Linguistics",
    doi = "10.18653/v1/2022.findings-naacl.35",
    pages = "465--475",
}

@incollection{SHNEIDERMAN2003364,
    title = {The Eyes Have It: A Task by Data Type Taxonomy for Information Visualizations},
    editor = {BENJAMIN B. BEDERSON and BEN SHNEIDERMAN},
    booktitle = {The Craft of Information Visualization},
    publisher = {Morgan Kaufmann},
    address = {San Francisco},
    pages = {364-371},
    year = {2003},
    series = {Interactive Technologies},
    isbn = {978-1-55860-915-0},
    doi = {https://doi.org/10.1016/B978-155860915-0/50046-9},
    url = {https://www.sciencedirect.com/science/article/pii/B9781558609150500469},
    author = {Ben Shneiderman},
}

@article{guan2025monitoring,
  title={Monitoring monitorability},
  author={Guan, Melody Y and Wang, Miles and Carroll, Micah and Dou, Zehao and Wei, Annie Y and Williams, Marcus and Arnav, Benjamin and Huizinga, Joost and Kivlichan, Ian and Glaese, Mia and others},
  journal={arXiv preprint arXiv:2512.18311},
  year={2025}
}

@article{chen2025towards,
  title={Towards reasoning era: A survey of long chain-of-thought for reasoning large language models},
  author={Chen, Qiguang and Qin, Libo and Liu, Jinhao and Peng, Dengyun and Guan, Jiannan and Wang, Peng and Hu, Mengkang and Zhou, Yuhang and Gao, Te and Che, Wanxiang},
  journal={arXiv preprint arXiv:2503.09567},
  year={2025}
}

@article{zhou2025improving,
  title={Improving human verification of llm reasoning through interactive explanation interfaces},
  author={Zhou, Runtao and Nguyen, Giang and Kharya, Nikita and Nguyen, Anh Totti and Agarwal, Chirag},
  journal={arXiv preprint arXiv:2510.22922},
  year={2025}
}

@misc{openai_api,
  title = {Reasoning models | OpenAI API},
  url= {https://developers.openai.com/api/docs/guides/reasoning},
  year = {2026},
  author = {OpenAI},
}

@misc{gemini_api,
  title = {Gemini thinking | Gemini API | Google AI for Developers},
  url = {https://ai.google.dev/gemini-api/docs/thinking},
  year = {2026},
  author = {Google}
}

@misc{claude_api,
  title = {Messages API | Anthropic},
  howpublished = {\url{https://platform.claude.com/docs/en/api/messages}},
  year = {2026},
  note = {Accessed: 2026-02-27},
  author = {Anthropic}
}

@misc{xai_inference_api,
  title = {Inference API - REST API Reference | xAI Docs},
  howpublished = {\url{https://docs.x.ai/developers/rest-api-reference/inference}},
  year = {2026},
  note = {Accessed: 2026-02-27},
  author = {xAI}
}

@misc{deepseek_api,
  title = {DeepSeek API | DeepSeek API Docs},
  howpublished = {\url{https://api-docs.deepseek.com/api/deepseek-api}},
  year = {2026},
  note = {Accessed: 2026-02-27},
  author = {DeepSeek}
}

@misc{qwen_api,
  title = {Qwen API Platform},
  howpublished = {\url{https://qwen.ai/apiplatform/}},
  year = {2026},
  note = {Accessed: 2026-02-27},
  author = {Qwen}
}

@misc{mistral_api,
  title = {API Specs | Mistral AI Docs},
  howpublished = {\url{https://docs.mistral.ai/api}},
  year = {2026},
  note = {Accessed: 2026-02-27}, 
  author = {Mistral AI}
}

@inproceedings{xie2024waitgpt,
  title={Waitgpt: Monitoring and steering conversational llm agent in data analysis with on-the-fly code visualization},
  author={Xie, Liwenhan and Zheng, Chengbo and Xia, Haijun and Qu, Huamin and Zhu-Tian, Chen},
  booktitle={Proceedings of the 37th Annual ACM Symposium on User Interface Software and Technology},
  pages={1--14},
  year={2024}
}

@inproceedings{lee2025llm,
  title={Llm attributor: Interactive visual attribution for llm generation},
  author={Lee, Seongmin and Wang, Zijie J and Chakravarthy, Aishwarya and Helbling, Alec and Peng, ShengYun and Phute, Mansi and Chau, Duen Horng Polo and Kahng, Minsuk},
  booktitle={Proceedings of the AAAI Conference on Artificial Intelligence},
  volume={39},
  number={28},
  pages={29655--29657},
  year={2025}
}

@article{cheng2025understanding,
  title={Understanding large language model behaviors through interactive counterfactual generation and analysis},
  author={Cheng, Furui and Zouhar, Vil{\'e}m and Chan, Robin Shing Moon and F{\"u}rst, Daniel and Strobelt, Hendrik and El-Assady, Mennatallah},
  journal={IEEE Transactions on Visualization and Computer Graphics},
  year={2025},
  publisher={IEEE}
}

@inproceedings{wang2025data,
  title={Data formulator 2: Iterative creation of data visualizations, with ai transforming data along the way},
  author={Wang, Chenglong and Lee, Bongshin and Drucker, Steven M and Marshall, Dan and Gao, Jianfeng},
  booktitle={Proceedings of the 2025 CHI Conference on Human Factors in Computing Systems},
  pages={1--17},
  year={2025}
}

@inproceedings{pluto,
author = {Srinivasan, Arjun and Setlur, Vidya and Satyanarayan, Arvind},
title = {Pluto: Authoring Semantically Aligned Text and Charts for Data-Driven Communication},
year = {2025},
isbn = {9798400713064},
publisher = {Association for Computing Machinery},
address = {New York, NY, USA},
url = {https://doi.org/10.1145/3708359.3712122},
doi = {10.1145/3708359.3712122},
booktitle = {Proceedings of the 30th International Conference on Intelligent User Interfaces},
pages = {1123–1140},
numpages = {18},
location = {
},
series = {IUI '25}
}

@inproceedings{kocielnik2019will,
  title={Will you accept an imperfect ai? exploring designs for adjusting end-user expectations of ai systems},
  author={Kocielnik, Rafal and Amershi, Saleema and Bennett, Paul N},
  booktitle={Proceedings of the 2019 CHI conference on human factors in computing systems},
  pages={1--14},
  year={2019}
}

\appendix
\clearpage
\newpage
\setcounter{table}{0}
\setcounter{figure}{0}

% \wy{Please start from a new page.}
\section{Exposure of Reasoning Traces}
\label{app:api-exposure}
\csw{This appendix summarizes how some major
LRM providers expose reasoning in user UIs and developer APIs. Table~\ref{tab:api-reasoning-exposure} shows details of 6 major providers (snapshot as of Mar.~2026).
% \wy{Given that the appendix is another independent file, the table ID should start from 1. So it should be Table 1. Same for other figures.}
}
\begin{table}[htbp]
\centering
\small
\begin{tabularx}{0.47\textwidth}{P{1.1cm} C{1.8cm} C{2cm} C{2cm}}

\hline
\hline
\textbf{Provider} & \textbf{Model (example)} & \textbf{API reasoning exposure} & \textbf{UI reasoning exposure} \\
\hline
\hline
OpenAI\cite{openai_api} & GPT-5 & summary & summary\\
\hline

Google\cite{gemini_api} & Gemini 3 & summary & summary\\
\hline

Anthropic\cite{claude_api} & Claude Opus 4  & raw reasoning &summary \\
\hline

xAI\cite{xai_inference_api} & Grok 3/4 & no &  summary \\
\hline

DeepSeek\cite{deepseek_api} & DeepSeek-R1 & raw reasoning & raw reasoning \\
\hline

Qwen\cite{qwen_api} & Qwen3/3.5 & raw reasoning & raw reasoning \\
\hline

Mistral\cite{mistral_api} & magistral & raw reasoning & raw reasoning  \\
\hline
\hline

\end{tabularx}
% \caption{Survey of how major LRM providers expose reasoning traces in user interfaces and developer-facing APIs, as of February 2026.}
\caption{A summary of 7 LRM providers in terms of the reasoning trace exposure in their user interfaces and APIs, as of Mar. 2026.}
\label{tab:api-reasoning-exposure}
\end{table}

\begin{table*}[htbp]
\centering
\small
\begin{tabular}{c c l l c c}
\hline
\textbf{ID} & \textbf{Age} & \textbf{Primary role(s)} & \textbf{Gender} & \textbf{LRM familiarity} & \textbf{Frequency of inspecting reasoning traces} \\
            &              & (select all that apply) &    & (scale 0--7) & (Never / Rarely / Sometimes / Often)\\
\hline
1 & 22 & Researcher (HCI/NLP/ML), Student & Male   & 7 & Sometimes \\
2 & 21 & Researcher (HCI/NLP/ML), Student & Male   & 6 & Sometimes \\
3 & 23 & Student                          & Female & 6 & Rarely    \\
4 & 26 & Researcher (HCI/NLP/ML), Engineer/Developer, Student & Female & 7 & Sometimes \\
5 & 29 & Researcher (HCI/NLP/ML)          & Male   & 7 & Often     \\
6 & 31 & Researcher (HCI/NLP/ML), Engineer/Developer, Student & Male   & 6 & Sometimes \\
7 & 23 & Engineer/Developer               & Female & 7 & Sometimes \\
8 & 26 & Engineer/Developer               & Male   & 7 & Sometimes \\
9 & 26 & Data Scientist                   & Male   & 7 & Often     \\
\hline
\end{tabular}
\caption{Participant details (age, background, gender, familiarity, and self-reported frequency of inspecting reasoning traces).}
\label{tab:user-details}
\end{table*}

\begin{figure}[ht]
    \centering
    \includegraphics[width=\linewidth]{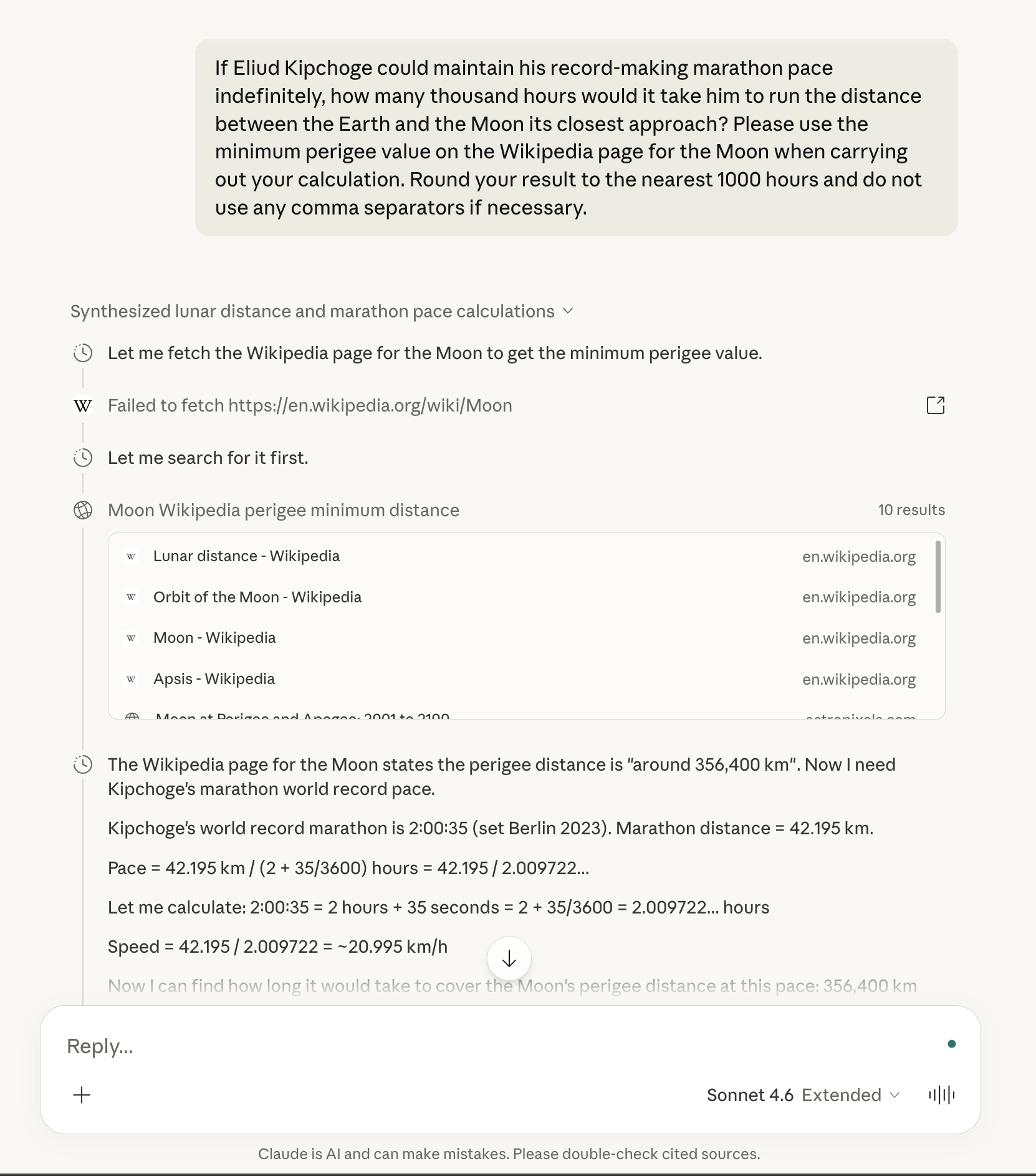}
    \caption{Example of Claude presenting a summarized reasoning trace with multiple steps.}
    \label{fig:claude}
\end{figure}

\begin{figure}[ht]
    \centering
    \includegraphics[width=\linewidth]{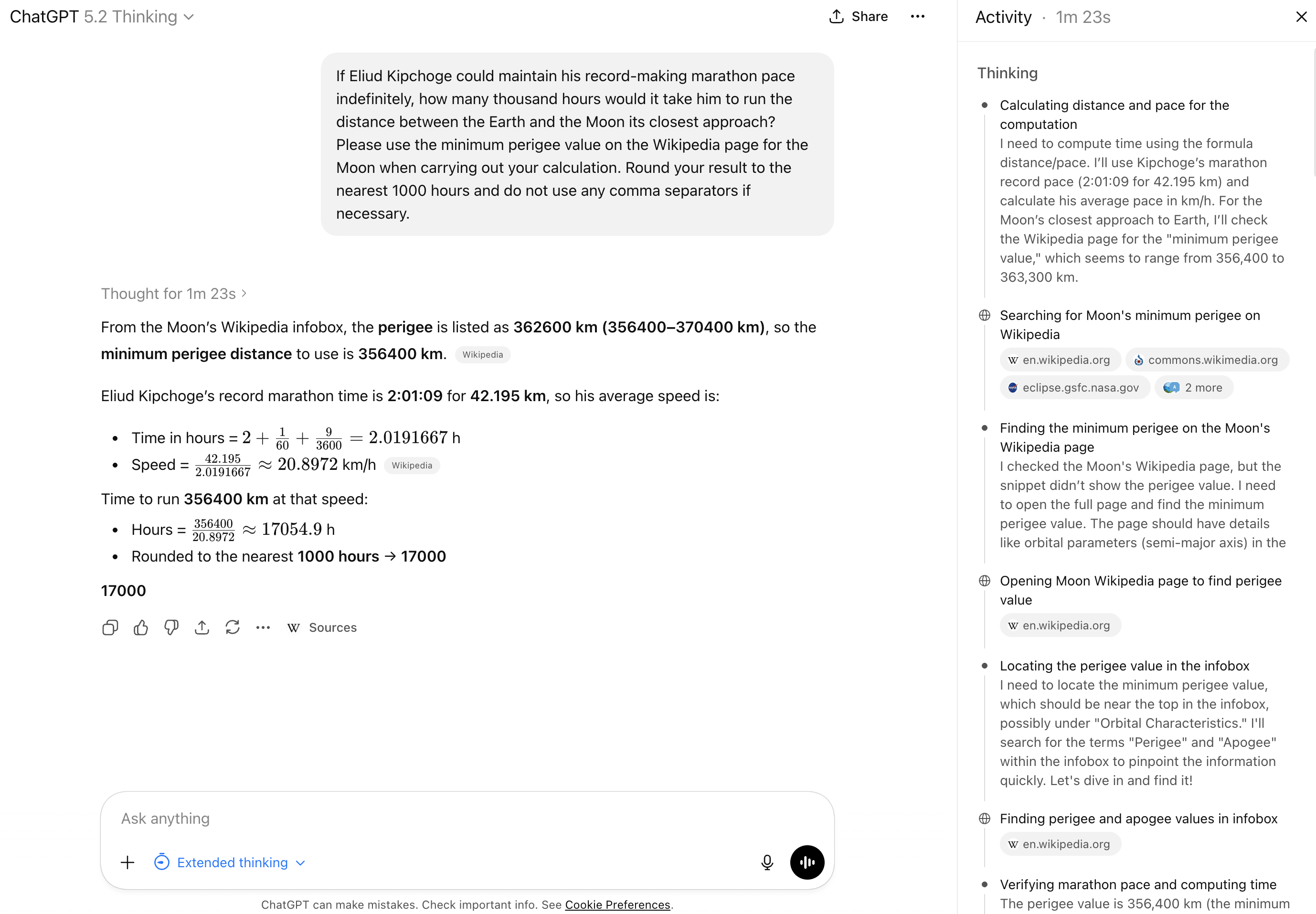}
    \caption{Example of ChatGPT presenting a summarized reasoning trace with brief explanations.}
    \label{fig:gpt}
\end{figure}

\csw{We use \textbf{summary} to mean provider-controlled, non-verbatim abstractions of internal reasoning (e.g.,~\autoref{fig:claude}~\autoref{fig:gpt}) that aim to increase transparency while reducing information overload and limiting sensitive-trace leakage. In contrast, \textbf{raw reasoning} denotes an explicit, model-generated reasoning trace returned as a first-class field or content block (e.g., \texttt{thinking} / \texttt{reasoning\_content}) that developers can programmatically access. If a provider exposes only \texttt{encrypted\_reasoning} (or an equivalent protected trace) that is not consumable by downstream systems, we treat it as no raw-reasoning access. Importantly, \name can still operate when \textbf{summary} contains dozens of steps (e.g., Claude~\autoref{fig:claude} ) or has corresponding explanations (e.g., ChatGPT~\autoref{fig:gpt} ), by performing step-level verification and dependency tracing over the exposed summarized text (albeit with reduced granularity compared to raw traces).}

\section{\textbf{Formative Study Details}}
\label{appendix:formative_study}
\csw{Table~\ref{tab:user-details} summarizes participants' demographics and prior exposure to large reasoning models (LRMs). Participants reported their age, gender, and primary role(s) (multi-select). We additionally collected a self-reported LRM familiarity rating on a 0--7 scale (0 = Never used; 7 = Use nearly every day) and how frequently they review reasoning traces during use (Never / Rarely / Sometimes / Often). These measures help contextualize participants' baseline experience and interpretation habits of LRM-generated reasoning.}

\section{\textbf{Technical Evaluation Result Analysis}}
\label{appendix:error_analysis}
\subsection{Metric Computation}
\csw{
We report precision, recall, and F1 at the \emph{per-sample} level: for each CoT instance, we first obtain $(P, R, F1, Acc)$. The overall results in Table~\ref{tab:results_comparison} are then reported as the \emph{macro-average} (arithmetic mean) of these per-sample metrics over all 13 samples.
}

\begin{table}[hbpt]
\centering
\small
\begin{tabular}{r c cc cc cc}
\hline
\textbf{ID} & \textbf{count} &
\multicolumn{2}{c}{\textbf{BIG-Bench}} &
\multicolumn{2}{c}{\textbf{\textit{Our}}} &
\multicolumn{2}{c}{\textbf{$\Delta$ (Our$-$BIG-Bench)}} \\
\cline{3-8}
 &  & \textbf{P} & \textbf{R} & \textbf{P} & \textbf{R} & \textbf{$\Delta$P} & \textbf{$\Delta$R} \\
\hline
1  & 91  & 0.82 & 0.69 & 0.75 & 0.69 & \DeltaBox{-0.07}{-0.07} & \DeltaBox{ 0.00}{ 0.00} \\
2  & 150 & 0.36 & 0.80 & 0.08 & 1.00 & \DeltaBox{-0.29}{-0.29} & \DeltaBox{ 0.20}{ 0.20} \\
3  & 76  & 0.25 & 0.73 & 0.25 & 0.73 & \DeltaBox{ 0.00}{ 0.00} & \DeltaBox{ 0.00}{ 0.00} \\
4  & 35  & 0.27 & 0.38 & 0.33 & 0.63 & \DeltaBox{ 0.06}{ 0.06} & \DeltaBox{ 0.25}{ 0.25} \\
5  & 59  & 0.76 & 0.92 & 0.79 & 0.92 & \DeltaBox{ 0.03}{ 0.03} & \DeltaBox{ 0.00}{ 0.00} \\
6  & 56  & 0.88 & 0.47 & 0.60 & 0.87 & \DeltaBox{-0.27}{-0.27} & \DeltaBox{ 0.40}{ 0.40} \\
7  & 70  & 0.17 & 1.00 & 0.06 & 0.75 & \DeltaBox{-0.11}{-0.11} & \DeltaBox{-0.25}{-0.25} \\
8  & 148 & 0.39 & 0.42 & 0.24 & 0.83 & \DeltaBox{-0.15}{-0.15} & \DeltaBox{ 0.42}{ 0.42} \\
9  & 28  & 0.40 & 0.25 & 0.35 & 0.63 & \DeltaBox{-0.04}{-0.04} & \DeltaBox{ 0.38}{ 0.38} \\
10 & 55  & 0.44 & 1.00 & 0.27 & 0.88 & \DeltaBox{-0.17}{-0.17} & \DeltaBox{-0.12}{-0.12} \\
11 & 117 & 0.09 & 0.33 & 0.13 & 1.00 & \DeltaBox{ 0.04}{ 0.04} & \DeltaBox{ 0.67}{ 0.67} \\
12 & 154 & 0.56 & 0.71 & 0.05 & 1.00 & \DeltaBox{-0.50}{-0.50} & \DeltaBox{ 0.29}{ 0.29} \\
13 & 132 & 0.21 & 0.86 & 0.07 & 0.50 & \DeltaBox{-0.14}{-0.14} & \DeltaBox{-0.36}{-0.36} \\
\hline
\end{tabular}
\caption{Per-sample comparison of  BIG-Bench vs Our Method. Each row corresponds to one evaluation sample, with its statement count and precision/recall. $\Delta$ denotes the per-sample difference (Our$-$BIG-Bench); cell background encodes $\Delta$ using a continuous diverging gradient.}
\label{tab:pr_recall_delta_continuous}
\end{table}
 
\subsection{Performance Analysis}
\csw{Table~\ref{tab:pr_recall_delta_continuous} reports per-sample precision and recall for the BIG-Bench prompt and our method on each CoT sample, together with the per-sample differences ($\Delta$ $=$ Our $-$ BIG-Bench). Overall, BIG-Bench achieves slightly higher precision in most samples (typically by $\sim$0.1), whereas our method yields higher recall in most samples (typically by $\sim$0.3). 
Notably, a few outliers (e.g., samples with IDs 2, 6, and 12) exhibit substantially larger precision gaps between the two methods; these outliers largely contribute to our lower mean F1 score. We analyze these samples in detail below.}
\begin{itemize}
    \item \csw{\textbf{Sample\#6.} For this sample, we observe a clear precision-recall trade-off: while our method yields lower precision (0.60), it achieves higher recall (0.87), ultimately resulting in a better overall F1 score (0.71) compared to BIG-Bench (0.60).}
    \item \note{\textbf{Sample\#2 \& Sample\#12. }For both samples, our detector achieved perfect recall (1.00), but precision was very low ($<0.1$). We found that this is mainly caused by an early \emph{core} statement being misclassified, which then triggered \emph{propagated} error labels for many downstream steps that were only conditionally dependent on that statement.\\
    In both cases, the detected errors appeared in long, contiguous bursts (Sample\#2: two bursts spanning 18 and 98 steps; Sample\#12: two bursts spanning 22 and 20 steps), suggesting a cascade pattern rather than many independent local mistakes. For instance, in Sample\#2, the burst of 18 contiguous flagged steps can be traced to a single core step stating \textit{``any sum of consecutive integers can be represented as a difference of two triangular numbers''}, which was marked as a factual error because the external verifier retrieved a related but non-equivalent claim as evidence and labeled it as ``refute''. This mismatch created a spurious contradiction signal at the core step and subsequently inflated propagated errors downstream.}
% \wy{It is very hard to understand these three examples. Perhaps we can show our visualization + original traces so that we can understand it?}
    
\end{itemize}
\csw{\textbf{Summary.} Across samples, the main performance difference comes from our recall-oriented design and its interaction with dependency propagation. While this improves coverage of erroneous steps, the dominant source of false positives is propagation: \textbf{44.48\%} of our total false positives are propagated errors. Outlier samples (e.g., \#2 and \#12) illustrate a typical failure mode where a single early core statement is incorrectly flagged, which then triggers long cascades of downstream flags. Importantly, these false positives are structured rather than scattered: \name helps users trace dense propagated error bursts back to a small set of core statements for human verification, which mitigates the practical cost of lower precision.}
% 我们发现我们的方法能够找到全部的错误(Recall = 1.00)，但由于存在将一个关键节点判断为 FactError，使得后续大量假设都被判定为错。
% 在这两个中我们首先关注到这些错误都非常连续 从Sample#2第46-65，96-193句都被判定为存在错误，Sample#12从第 7-28, 198-217 步骤都被判定为存在错误。观察具体信息我们发现这些 error 是从某些早期的核心论点被判定为 error propagated 得到：

%对于，例如“any sum of consecutive integers can be represented as a difference of two triangular numbers”被判定为错由于其从 Quora 找到了证明 “there is no triangular number that is equal to the sum of two consecutive positive integers”. 一些statement被判断为logic error 是由于存在 “that means that \( \frac{m-1}{2} \) must be a half-integer if \( m \) is odd, or an integer if \( m \) is even.”on mixed-content steps: a single sentence may contain multiple sub-claims, where one sub-claim is incorrect but another is correct and later reused by downstream steps. For example, a step might state ``\textit{The Hubble Space Telescope was launched in 1990 by the European Space Agency.}'': contains a correct sub-claim (launched in 1990) and an incorrect one (launched by ESA); later steps may rely only on the correct launch year (e.g., to compute years of operation), so our step-level flagging of the entire sentence is counted as a false positive, reducing precision.

\section{\textbf{Prompting Details for \name}}
\label{appendix:prompting_details}
 \csw{The backend model is configurable and can be replaced without changing the pipeline logic, subject to practical trade-offs (cost/latency/performance).
 \subsection{Error Detection Pipeline}
 For Step Classification and Premise Tree Generation, we adopt the prompt design from ThoughtAnchor~\cite{bogdan2025thought} and PARC~\cite{mukherjee2025premise}, and have small modifications like excluding unnecessary output or adjusting input/output format according to our needs.}
% \clearpage
% \onecolumn
\begin{tcolorbox}[
  title=Step Classification,
  colback=white, colframe=gray,
  breakable,
  enhanced,
  listing options={breaklines=true, columns=fullflexible, basicstyle=\ttfamily\footnotesize},
]
\textbf{Purpose:} Assign function tag to each step according it's function role for quick filtering.\\
\textbf{Inputs:} \{TASK\_QUESTION\}, \{FULL\_COT\_STEP\} \\
\textbf{Model/Params:} gpt-5, max\_tokens=10000 \\
\textbf{Output Schema:} JSON with fields \{step\_index, function\_tag\}

\tcblower
% \begin{verbatim}
\ttfamily\small
You are an expert in interpreting how language models solve math problems using multi-step reasoning. Your task is to analyze a Chain-of-Thought (CoT) reasoning trace, broken into discrete steps, and label each step with:

**function\_tag**: One or more labels that describe what this step is *doing* functionally in the reasoning process.

---

\#\#\# Function Tags:

1. `problem\_setup`: 
    Parsing or rephrasing the problem (initial reading or comprehension).
    
2. `plan\_generation`: 
    Stating or deciding on a plan of action (often meta-reasoning).
    
3. `fact\_retrieval`: 
    Recalling facts, formulas, problem details (without immediate computation).
    
4. `active\_computation`: 
    Performing algebra, calculations, manipulations toward the answer.
    
5. `result\_consolidation`: 
    Aggregating intermediate results, summarizing, or preparing the final answer.
    
6. `uncertainty\_management`: 
    Expressing confusion, re-evaluating, proposing alternative plans (includes backtracking).
    
7. `final\_answer\_emission`: 
    Explicit statement of the final boxed answer or earlier steps that contain the final answer.
    
8. `self\_checking`: 
    Verifying previous steps, Pythagorean checking, re-confirmations.

9. `unknown`: 
    Use only if the step does not fit any of the above tags or is purely stylistic or semantic.

---

\#\#\# Output Format:

Return a single dictionary with one entry per step, where each entry has:\\
- the step index (as the key, converted to a string),\\
- a dictionary with: `"function\_tag"`: list of tag strings\\
    
Here's the expected format:
\begin{verbatim}
```language=json
{{
    "<step_index>": {{
    "function_tag": ["..."],
    ...
}}
```
\end{verbatim}

Here is the problem:\\
\{TASK\_QUESTION\}\\

Here is the full Chain of Thought, broken into steps:\\
\{FULL\_COT\_STEP\}\\

Now label each step with function tags.
\end{tcolorbox}

% \subsection{Verifiability Assessment}
\begin{tcolorbox}[
  title=Verifiability Assessment,
  colback=white, colframe=gray,
  breakable,
  enhanced,
  listing options={breaklines=true, columns=fullflexible, basicstyle=\ttfamily\footnotesize},
]
\textbf{Purpose:} Determine whether a CoT step contains an objectively checkable claim (via external facts or logical validation).\\
\textbf{Inputs:} \{FULL\_COT\_STEP\} \\
\textbf{Model/Params:} gpt-5, max\_tokens=10000 \\
\textbf{Output Schema:} JSON with fields \{step\_index, category, explanation, confidence\}

\tcblower
\ttfamily\small
[SYSTEM]
You are an expert reasoning analyst.

Goal\\
For EACH input statement, output ONE of two categories:\\
 - Verifiable: contains at least one objectively checkable claim, confirmable via external knowledge sources or logical validation.\\
 - Non\_verifiable: contains no objectively checkable claim (e.g., instructions, planning/organization, opinions, hedging, or reflective/meta commentary).\\

Decision checklist (for INTERNAL deliberation only; DO NOT output these steps)\\
1) Checkability test: Does the statement assert a claim that could be objectively verified as true/false using external sources or formal/logical validation?\\
2) Evidence availability: In principle, could a verifier consult public knowledge, data, or compute a proof/check (without relying on extra unstated assumptions)?\\
3) Non-claim filter: If the statement is purely procedural, subjective, or meta/organizational (no factual or logically testable claim), label it Non\_verifiable.\\

Output Schema (JSON)
\begin{verbatim}
[
  {
    "id": "<step_index>",
    "category": "Verifiable|Non_verifiable",
    "explanation": "<string>",
    "confidence": "<number 0..1>"
  } 
]
\end{verbatim}

Example (ILLUSTRATIVE)
[AN EXAMPLE HERE, Omitted due to space]

[USER]\\
You are given a batch of statements in JSON. Classify each into Verifiable|Non\_verifiable using the system rules. Output only the results JSON.\\

INPUT:\\
\{FULL\_COT\_STEP\}\\

RESPONSE (STRICT JSON ONLY):

\end{tcolorbox}

\begin{tcolorbox}[
  title=Premise Tree Construction,
  colback=white, colframe=gray,
  breakable,
  enhanced,
  listing options={breaklines=true, columns=fullflexible, basicstyle=\ttfamily\footnotesize},
]
\textbf{Purpose:} For each verifiable step, retrieve the minimal set of prior verifiable steps that serve as its premises, given the preceding reasoning context.\\
\textbf{Inputs:} \{TASK\_QUESTION\}, \{COT\_STEP(Verifiable)\}, \{COT\_CONTEXT\} \\
\textbf{Model/Params:} gpt-5, max\_tokens=10000 \\
\textbf{Output Schema:} JSON with fields \{step\_index, function\_tag\}

\tcblower
\ttfamily\small
You are provided with a question, a partial solution, and the next step in the solution. Your task is to identify the steps that serve as premises for the given next step.
A step qualifies as a premise if the next step directly relies on information from that step. Based on the identified premises, the correctness of the next step should be fully verifiable.\\

Question (Step 0):\\
\{TASK\_QUESTION\}\\

Solution so far:\\
\{COT\_CONTEXT\}\\

Next step to analyze:\\
\{COT\_STEP\}\\

For the step above, identify which previous steps (including Step 0 - the question) are premises and explain why each one is necessary. Remember:\\
1. A step cannot be a premise to itself\\
2. The question (Step 0) can be a premise if used directly\\

Generate **ONLY** the premises and nothing else.\\
Format your response with one premise per line as:\\
Step X: [explanation of why this step is necessary for the current step]\\ \{fewshot\_template\}

\end{tcolorbox}

\begin{tcolorbox}[
  title=NL to Symbolic,
  colback=white, colframe=gray,
  breakable,
  enhanced,
  listing options={breaklines=true, columns=fullflexible, basicstyle=\ttfamily\footnotesize},
]
\textbf{Purpose:} Convert premises and conclusion from natural language into symbolic formulas.\\
\textbf{Inputs:} \{target\_statement\}, \{related\_statements\}, \{full\_reasoning\}, \{question\_context\}, \{declarations\_and\_constraints\} \\
\textbf{Model/Params:} gpt-5, max\_tokens=10000 \\
\textbf{Output Schema:} JSON with fields \{declarations, constraints, statements, target\_statement\} where target\_statement=\{sentence, FL\}.\\
\tcblower
\ttfamily\small

You are an expert in formal logic and automated reasoning.
You are given:\\
1. "target\_statement": The target statement to be transformed to formal logic (in natural language).\\
2. "related\_": A list of supporting statements (in natural language) relevant to the main statement that need to be transformed to formal logic as constraints.\\
3. "full\_reasoning": The complete reasoning chain in natural language for refining the related statements.\\
4. "question\_context": The original question text for background.\\
5. "declarations \& constraints": Logic declarations and variable domains in the question, as would be used in formal logic (e.g., function, variable, and domain definitions). Any extra constraints or axioms given or derived from the question.\\

Your task:\\
- Convert the "target\_statement" (in natural language) into formal logic.  \\
- Convert every "related\_statement" into a constraint in formal logic. \\ 
- Extract additional valid logical statements from "full\_reasoning" that are not simple restatements of the related\_statements, and put them into "statements".  \\
- Preserve all declarations and constraints from the input.  \\
- Use the exact variable names and function names from the provided declarations.  \\
- Rewrite each of them strictly using the same syntax, function names, and variable names provided in the declarations and constraints.  \\
- Do not summarize, explain, or add extra reasoning — only produce logical statements.  \\
- If a sentence in the "full\_reasoning" implies another statement, express that with `Implies(...)`.  \\
- If a sentence negates a statement, use `Not(...)`.  \\
- Maintain the given naming conventions.   \\
- Do not repeat the original problem constraints; only translate the "full\_reasoning" into logical form.\\

Output Format:\\
Return a JSON object with the following fields:\\
1. "declarations": An array of the formal logic declarations from related\_statements and the given declarations(as code or formal expressions).\\
2. "constraints": An array of the formal logic constraints for related\_statements (as code or formal expressions).\\
3. "target\_statement":\\
  An object with:\\
  - sentence: The target statement (in natural language).\\
  - FL: The formal logic representation.\\

\end{tcolorbox}

\begin{tcolorbox}[
  title=Symbolic to Z3 Code,
  colback=white, colframe=gray,
  breakable,
  enhanced,
  listing options={breaklines=true, columns=fullflexible, basicstyle=\ttfamily\footnotesize},
]
\textbf{Purpose:} Convert symbolic formulas into executable Z3 (Python) code for verification.\\
\textbf{Inputs:} \{declarations\}, \{constraints\}, \{target\_statement\} \\
\textbf{Model/Params:} gpt-5, max\_tokens=10000 \\
\textbf{Output Contract:} Return only valid Python code using \texttt{z3py}. The code must define (1) all declarations, (2) \texttt{constraints} = [...], and (3) \texttt{target\_statement} = [...].\\

\tcblower
\ttfamily\small
You are given 3 inputs: "declarations", "constraints", and "target\_statement".\\
Your task:\\
- Convert these inputs into valid Z3 Python code.\\
- The code must include:\\
  1. Declarations (using EnumSort, Function, Const, etc.).\\
  2. A list named `constraints` containing the constraints.\\
  3. A list named `target\_statement` containing the target\_statement logical statements.\\
- Ensure all variable names and function names exactly match the ones in the declarations and constraints.\\
- Use `Const` for quantified variables.\\
- Do not include solver code, explanations, or any extra strings. Only return the pure Python code such that it can be executed directly.\\
---
\end{tcolorbox}

\subsection{\name Hierarchical Summary Construction}
% \wy{Interface? We are not introducing interface here, isn't it?}
The prompt is designed generate summaries to consider both the global reasoning trajectory of the entire CoT and the functional roles of individual steps, ensuring that the resulting hierarchy accurately reflects the actual reasoning process.

\begin{tcolorbox}[
  title=CoT Section Structuring (Hierarchy + Summaries),
  colback=white, colframe=gray,
  breakable,
  enhanced,
  listing options={breaklines=true, columns=fullflexible, basicstyle=\ttfamily\footnotesize},
]
\textbf{Purpose:} Assign each reasoning step a hierarchical plan structure and generate an abstract intent description.\\
\textbf{Inputs:} \{FULL\_COT\_STEPS\} (ordered list of steps with indices and step types)\\
\textbf{Model/Params:} gpt-5, max\_tokens=10000 \\
\textbf{Output Contract:} Output \textbf{only} a JSON object matching the schema below. Do not include explanations or extra text.\\

\tcblower
\ttfamily\small
\begin{verbatim}
You are an expert in analyzing Chain-of-Thought (CoT) reasoning traces. Your goal is to recover the *actual reasoning process* by jointly modeling:
(1) the global reasoning trajectory of the entire CoT (major phases and transitions), and
(2) the functional roles of individual sentences (via the provided function tags).

You will be given an input JSON object. Each entry represents one sentence-level reasoning step with:
- "sentence": the step text
- "function_tag": a tag from the tag set below

### Function Tags:
1. `problem_setup`: 
    Parsing or rephrasing the problem (initial reading or comprehension).
2. `plan_generation`: 
    Stating or deciding on a plan of action (often meta-reasoning).
3. `fact_retrieval`: 
    Recalling facts, formulas, problem details (without immediate computation).
4. `active_computation`: 
    Performing algebra, calculations, manipulations toward the answer.
5. `result_consolidation`: 
    Aggregating intermediate results, summarizing, or preparing the final answer.
6. `uncertainty_management`: 
    Expressing confusion, re-evaluating, proposing alternative plans (includes backtracking).
7. `final_answer_emission`: 
    Explicit statement of the final boxed answer or earlier steps that contain the final answer.
8. `self_checking`: 
    Verifying previous steps, Pythagorean checking, re-confirmations.

### Task
Construct a *hierarchical plan outline* over the CoT steps. The hierarchy must reflect:
- the *global trajectory*: how the reasoning moves from setup → plan → retrieval/computation → consolidation → checking → finalization,
- and the *local roles*: why each sentence exists (its function tag) and how it supports a higher-level intent.

### Output Format (JSON only; no prose)
Output a JSON object keyed by sentence indices (as strings). Each key corresponds to one sentence index from the input. Each entry MUST follow the following schema:
{
  "function_tag": "...", 
  "depth": <int>, // 0 for top-level, child depth = parent depth + 1 (max 2)
  "abstract": "<short intent phrase 2-5 words>"
}

### Output Example (illustrative only)
{
    "4": {
        "function_tag": "plan_generation",
        "depth": 0,
        "abstract": "Define approach"
    },
    "24": {
        "function_tag": "uncertainty_ management"
        "depth": 0,
        "abstract": "Found potential errors"
    },
    "34": {
        "function_tag": "self_checking",
        "depth": 1,
        "abstract": "Correct errors"
    },
}

Input JSON:
{{FULL_COT_STEPS\}}

Process the input now and provide only the output JSON.
---
\end{verbatim}
\end{tcolorbox}

\clearpage
\twocolumn

\end{document}